\RequirePackage{amsthm}
\documentclass[sn-mathphys,Numbered]{sn-jnl}

\usepackage{graphicx}%
\usepackage{multirow}%
\usepackage{amsmath,amssymb,amsfonts}%
\usepackage{amsthm}%
\usepackage{mathrsfs}%
\usepackage[title]{appendix}%
\usepackage{xcolor}%
\usepackage{textcomp}%
\usepackage{manyfoot}%
\usepackage{booktabs}%
\usepackage{algorithm}%
\usepackage{algorithmicx}%
\usepackage{algpseudocode}%
\usepackage{listings}%

\usepackage{hyperref}   
\hypersetup{
	colorlinks=true,
	linkcolor=blue,
	citecolor=blue,
	urlcolor=blue,
}

\usepackage{mathtools}

\usepackage{lipsum}
\usepackage{diagbox}
\usepackage{indentfirst}

\usepackage{tabularray}
\usepackage{tabularx}
\usepackage{multirow}
\usepackage{geometry}
\usepackage{hhline}
\usepackage{makecell}
\DeclarePairedDelimiterX\set[1]\{\}{\nonscript\,#1\nonscript\,}
\usepackage{cleveref}
\usepackage[parfill]{parskip}
\usepackage{threeparttable} 
\usepackage[numbers]{natbib}
\usepackage{kotex}

\theoremstyle{thmstyleone}%
\theoremstyle{thmstyletwo}%

\theoremstyle{thmstylethree}%

\begin{document}

\title[Article Title]{Vehicle Suspension Recommendation System: Multi-Fidelity Neural Network-based Mechanism Design Optimization}



\author[1]{Sumin Lee}\email{smlee1996@kaist.ac.kr}

\author*[2]{Namwoo Kang}\email{nwkang@kaist.ac.kr}

\affil[1]{Department of Mechanical Engineering, Korea Advanced Institute of Science and Technology (KAIST), 291 Daehak-ro, Yuseong-gu, Daejeon 34141, Republic of Korea}

\affil[2]{Cho Chun Shik Graduate School of Mobility, Korea Advanced Institute of Science and Technology (KAIST), 193 Munji-ro, Yuseong-gu, Daejeon 34051, Republic of Korea}


\abstract{
Mechanical mechanisms are designed to perform specific functions in a variety of fields. In most cases, there is not a unique mechanism that performs a well-defined function. For example, vehicle suspensions are designed to improve driving performance and ride comfort, but different types are available depending on the environment in which they are used. This variability in design due to different usage environments makes performance comparison difficult. In addition, the industry's traditional design process is multi-step, gradually reducing the number of design candidates while performing costly analysis to achieve target performances. Recently, artificial intelligence models have been used to replace the computational cost of finite element analysis (FEA). However, there are limitations in data availability and different analysis environments, especially when moving from low-fidelity to high-fidelity analysis. In this paper, we propose a multi-fidelity design framework aimed at recommending optimal types and designs of mechanical mechanisms. As an application, vehicle suspension systems were selected and several types were defined. For each type, mechanism parameters were generated and converted into 3D CAD models, followed by low-fidelity rigid body dynamic analysis under driving conditions. To effectively build a deep learning-based multi-fidelity surrogate model, the results of the low-fidelity analysis were analyzed using Density-Based Spatial Clustering of Applications with Noise (DBSCAN) and sampled at 5\% for the high-cost flexible body dynamic analysis. After training a multi-fidelity model, a multi-objective optimization problem was formulated for the performance metrics of each suspension type. Finally, we recommend the optimal type and design based on the input (sprung mass) to optimize the ride comfort-related performance metrics. Finally, to validate the proposed methodology, we extracted basic design rules of Pareto solutions using data mining techniques. We also verified the effectiveness and applicability by comparing the results with those obtained from a conventional deep learning-based design process.}


\maketitle
\newpage
\section{Introduction}\label{sec1}
\subsection{Background}
The purpose of a mechanical mechanism is to perform a specific task. However, there can often be multiple mechanisms that can achieve the same functionality(\citeauthor{mcgovern1973kinematic, xu2024tracked}). A well-known example in the field of industry is that of vehicle suspension systems(\citeauthor{jiregna2020review}). The objective of vehicle suspension systems is to enhance ride quality and handling. These systems are available in various configurations, including MacPherson struts, double wishbone, and multi-link suspension systems. The selection process is inherently complex due to each type's unique advantages and challenges(\citeauthor{jiregna2020review, sharp1987road}).

Recently, the automotive industry has increasingly relied on computational methods to enhance suspension design processes(\citeauthor{koulocheris2017comparison,reddy2016comprehensive, fujita1998design}). Traditional design approaches have typically involved sampling specific design solutions within the design space, performing analyses on these discrete samples, and then progressing to more detailed designs based on the most promising samples. However, this method has a significant drawback: it only evaluates the potential of a given design space through its discrete samples, thus failing to fully exploit the potential within the entire design space.

With the advent of artificial intelligence, researchers sought to overcome these limitations by using predictive models such as deep-learning to select optimal values from continuous design variables(\citeauthor{kim2022deep, panchal2019machine, patel2021artificial, ramu2022survey}). This innovation aimed to streamline the design process and maximize the potential of the design space. Nonetheless, as product development progresses, the analysis environment typically transitions from low- to high-fidelity simulations, reducing the number of samples available for high-fidelity analysis. Despite constructing a robust learning model using low fidelity analysis data, there is still uncertainty about performance metrics in high-fidelity analyses due to potential discrepancies between low-fidelity and high-fidelity results. For example, assumptions applied in the low-fidelity analysis (e.g., rigid body analysis, ignoring friction) could yield different performance metrics under the same conditions in the high-fidelity analysis. Such differences, if considerable, can lead to critical issues that cannot be ignored. In addition, selecting the most appropriate suspension type and design parameters remains challenging, given the need to balance multiple objectives, such as minimizing suspension travel, acceleration and maximizing durability.

\subsection{Research goal}
In this paper, to effectively improve the mentioned challenges, we propose a multi-fidelity design framework aimed at recommending optimal types and designs of mechanical mechanisms, specifically vehicle suspension systems. Our approach integrates low-fidelity (rigid-body dynamics) analyses with high fidelity (multi flexible-body dynamics) simulations, using deep learning-based surrogate models to bridge the gap between fidelity levels. The multi-fidelity approach balances computational efficiency and prediction performance metrics by combining low-fidelity, cost-effective simulations with high-fidelity, accurate simulations. This ensures that the potential within the design space is significantly utilized and that the limitations of both traditional and AI-based design methods are significantly improved. Using advanced data mining techniques and multi-objective optimization, we aim to provide designers with robust guidelines and optimized solutions for suspension design.

In vehicle suspensions, specific performance metrics can be optimized for minimization or maximization, such as minimizing suspension travel, acceleration, or maximizing durability. However, some performance metrics, such as first mode frequency, cannot simply be optimized for higher or lower values but must be referenced to ensure that detailed design specifications are satisfied. This distinction is critical to achieving balanced and effective suspension designs. In particular, our framework enables designers to input the sprung mass of a vehicle and generate optimal design solutions (Pareto fronts) for several suspension types based on different first-mode frequencies. Designers can select the most appropriate design configuration from these Pareto optimal solutions. We will demonstrate the effectiveness of our framework through a case study on three types of suspensions, including MacPherson strut and double wishbone (low \& high mount) suspensions. The results will demonstrate the potential of our approach to increase design efficiency, reduce costs, and improve overall vehicle performance. The main contributions of this framework can be summarized as follows:

\begin{enumerate}
\item To the best of our knowledge, this is the first study to integrate traditional and AI-based design methods with a multi-fidelity framework for vehicle suspension mechanism optimization.

\item Proposed deep learning-based multi-fidelity framework that combines low-fidelity rigid-body dynamic analyses with high-fidelity flexible-body dynamic analysis, addressing the limitations of discrete sample evaluations in traditional design methods.

\item The framework enables effective optimization of multiple vehicle suspension types, including MacPherson strut and double wishbone (low \& high mount) suspensions, through advanced data mining and multi-objective optimization techniques.

\item Generate Pareto optimal design solutions based on various first mode frequencies, allowing designers to select the most appropriate suspension configuration for specific vehicle requirements.

\item Provide practical design guidelines by extracting fundamental design rules from Pareto solutions using data mining techniques, aiding in the future design of vehicle suspension systems.

\item The proposed methodology is validated through a comprehensive case study on three types of suspension systems, showcasing its potential to enhance design efficiency, reduce costs, and improve overall vehicle performance.
\end{enumerate}

This paper is organized as follows. Section \ref{sec2} summarizes the related studies, and Section \ref{sec3} covers the overall framework introduction and methodology for each stage. Section \ref{sec4} includes discussions by extracting the application results and design rules of the proposed framework, and compares the proposed framework with the conventional AI-based approach applied to suspension design. Finally, Section \ref{sec5} includes conclusions, limitations, and future work.

\section{Related works}\label{sec2}
In recent years, the industrial field has increasingly turned to artificial intelligence (AI) models, particularly deep learning, to address complex engineering challenges. Deep learning models can effectively utilize vast amounts of data continuously accumulated in the industry(\citeauthor{ullman1988model, krahe2022ai, shan2010survey}), such as automotive systems(\citeauthor{kang2014solving, shin2023wheel, grigorescu2020survey,gobbi2001analytical,alkhatib2004optimal,abdelkareem2018energy}), automated robotic manipulators(\citeauthor{wang2020deep, purwar2023deep, soori2023artificial}), smart factories(\citeauthor{chen2017smart, kotsiopoulos2021machine}), and beyond(\citeauthor{panchal2019machine, patel2021artificial}). Specifically, it has the potential to significantly enhance computational efficiency and accuracy in processes that traditionally rely on resource-intensive simulations like finite element analysis (FEA) and computational fluid dynamics (CFD)(\citeauthor{zhao2021adaptive, phiboon2021experiment, kennedy2000predicting}). The trend toward AI-driven solutions has led to various studies of optimization or methodologies to take advantage of the engineering field.
One noted field of research within this broader trend is the application of deep learning techniques to the design and optimization of mechanical mechanisms. \citeauthor{xue2023multi} explored a data-driven approach that maximizes computational efficiency by offering complex dynamic analysis solutions, extending the application of deep learning-based solvers to predict smooth solutions. This advancement is particularly crucial in scenarios where traditional methods struggle with the computational demands of real-time or high-fidelity simulations. Additionally, \citeauthor{go2024efficient} introduced an efficient method combining deep learning models with principal component analysis (PCA). This method leverages PCA to reduce the dimension of large datasets, enabling faster training times without compromising accuracy. \citeauthor{chen2015multi} conducted a study using kriging surrogate models for multi-objective optimization to improve vehicle ride comfort. This study bears similarities to the multi-objective optimization methods discussed in this paper, particularly in its goal of elucidating the relationship between design parameters and ride comfort. However, our proposed method differs by addressing scenarios where design decisions must be based solely on reference performance and simultaneously considering various suspension mechanism types. This distinction allows for a broader design space exploration, thereby enhancing the potential to obtain optimal performance.

Despite these advances, the conventional product design process typically involves multiple stages or iterations, each requiring different analysis conditions(\citeauthor{yoo2021integrating, lei2022ai, schlicht2021integrative}). The early stages of the design process often focus on prototype exploration, emphasizing simplified modeling and functionality. In contrast, later stages involve conducting high-fidelity simulations (such as FEA or CFD) on designs derived from earlier explorations to optimize and reduce the candidate design space. However, this traditional approach is not without its limitations.
First, there is a significant mismatch between the analysis environments and the types of performance metrics considered at each stage. For example, rigid body dynamics analysis does not account for deformation, making it impossible to calculate maximum stress values. On the other hand, transient dynamics analysis partially incorporates finite element methods, allowing for the calculation of maximum stress. This fundamental difference underscores the challenge of aligning performance metrics across different stages of the design process. Second, as the process progresses to stages requiring high-fidelity simulations, the number of explored designs and candidate solutions tends to decrease exponentially. This presents a significant challenge, particularly given the limited availability of sufficient datasets for the effective application of deep learning methods. In order to address the issues of data scarcity and high computational costs prevalent in the high-fidelity simulation stages, the concept of multi-fidelity has been proposed(\citeauthor{forrester2007multi, han2012hierarchical, peherstorfer2018survey}). The multi-fidelity approach aims to balance the accuracy of models derived from low- and high-fidelity simulations, providing a more efficient framework for optimization across various stages of the design process.

Multi-fidelity models have also been found to be applicable in flexible multibody dynamics analysis, where they address the computational challenges posed by including nonlinearities in dynamic simulations. \citeauthor{han2021dnn} proposed an algorithm for efficiently training a two-stage deep neural network designed for dynamic simulations that require nonlinear considerations. This approach targets the substantial computational costs of transient dynamic analysis including flexible bodies. Additionally, \citeauthor{xue2023multi} utilized analytical models of suspension systems and multibody models as low- and high-fidelity models to solve optimization problems related to filtering vibrations from road surfaces. While this approach is well-suited for guiding exploration during the early stages of the design process, it is essential to note that simplified models may not fully capture the complex nonlinear geometries inherent in multiple mechanisms.

Most cases mentioned above focus on optimization problems within the design process. However, in real-world design process scenarios, there are instances where design decisions must be referenced by performance values that cannot be optimized(\citeauthor{gomes2016multi, beck1999multi, privitera2017human, xu2022integrated, zhang2023serial, xie2013multi}). The main application proposed for vehicle suspension in this paper is that the natural frequency of a vehicle suspension system must be selected based on the vehicle's intended use and driving conditions, ensuring it remains within a specific range. Various factors, such as the joint position in the mechanism, sprung mass, and the mechanical properties of the shock absorber, influence the natural frequency of a suspension system. However, discrepancies in natural frequency values can arise between low-fidelity analyses (rigid body dynamics) and high-fidelity analyses (transient dynamics), underscoring the need for a multi-fidelity framework. Moreover, the optimal design solutions corresponding to the selected natural frequency may differ, necessitating real-time analysis and recommendation of optimal design distributions for the specified frequency. This study addresses the challenge of optimizing design solutions while improving data scarcity and performance inconsistency between high and low-fidelity simulations.

\section{Multi-fidelity suspension design optimization framework}\label{sec3}
\begin{figure*}[thb]
\centering
 \includegraphics[width=1\textwidth]{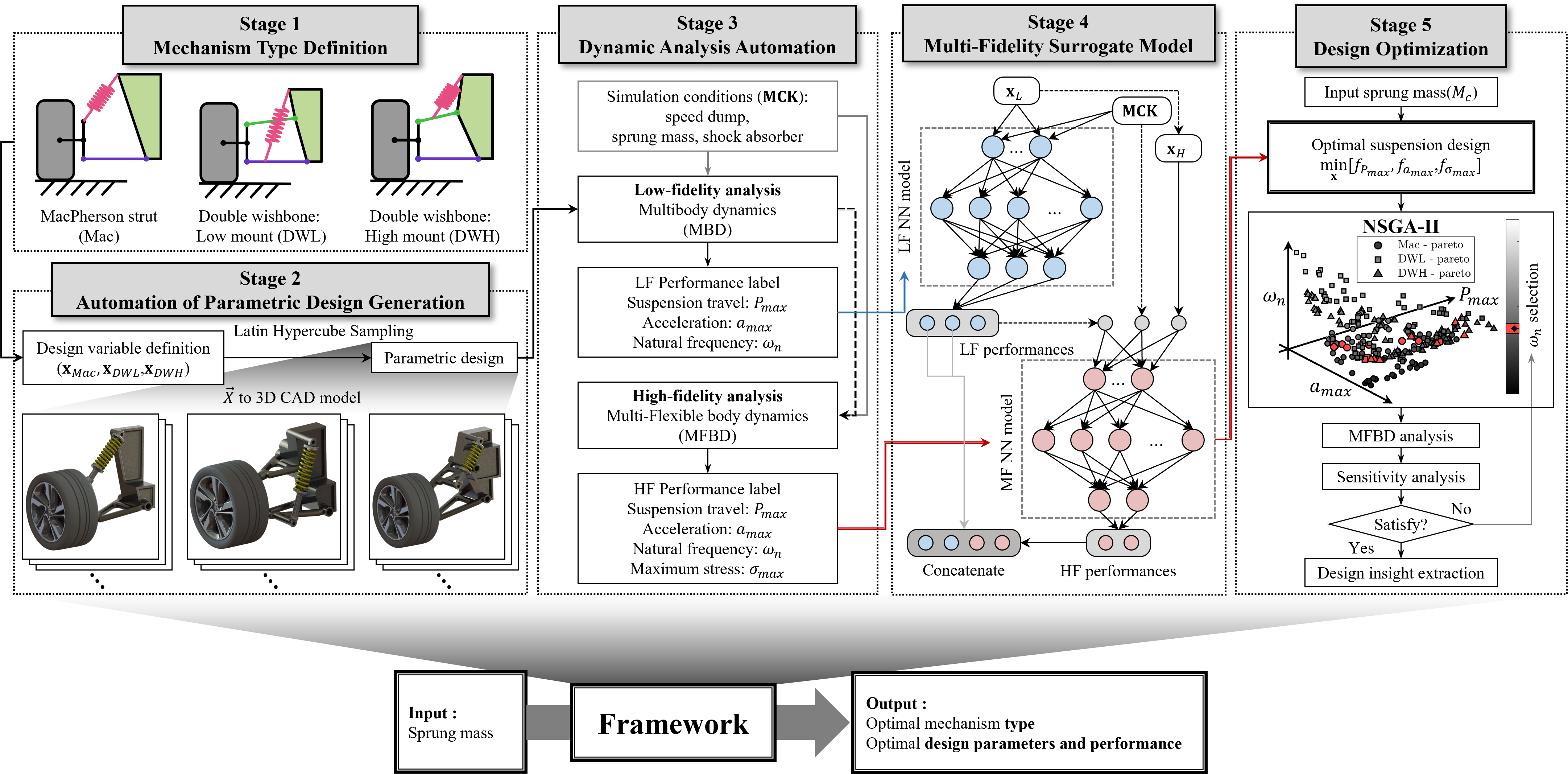}
\caption{Multi-fidelity suspension design optimization framework}
\label{fig:Framework}
\vspace{-9pt}
\end{figure*}

The deep-learning based multi-fidelity mechanism design framework proposed in this study consists of five stages, and its overall flowchart is shown in Fig. \ref{fig:Framework}. First of all, three types of suspensions are defined, and they are sampled by the functions that calculate the joints of the parametric-based mechanisms (Stage 1). Then, the design sample in the design space is converted into a 3D CAD model through parametric design (Stage 2). Based on the joint and geometric conditions calculated in Stage 1, an assembling for automatic dynamic analysis is established and used for low-fidelity and high-fidelity analysis (Stage 3). To predict the performance metrics of each suspension type, we train a multi-fidelity model based on the data performed in the low-fidelity and high-fidelity analysis (Stage 4). Finally, the already trained multi-fidelity surrogate model recommends optimal types and parameters via the non-dominated sorting genetic algorithm-II (NSGA-II) (Stage 5). The following subsections are organized to detail each stage of the proposed framework.

\subsection{Stage 1: Mechanism type definition}\label{subsec3_1}
In Stage 1, several mechanism types of the suspension system selected for the proposed application of the framework are defined. A suspension system's primary function is to mitigate vibrations experienced during road driving, thereby enhancing vehicle stability and ride comfort. Consequently, the suspension must provide degrees of freedom between the wheel and the vehicle body, with shock absorbers to effectively absorb vibrations.
Suspension systems enable relative motion between the wheel and the vehicle body while damping vibrations. This is primarily achieved through springs and shock absorbers, with the configuration varying depending on the type of suspension mechanism.
This study focuses on three representative types of suspension systems as shown in Fig. \ref{fig:mechanism type definition}: MacPherson strut, double wishbone (low mount), and double wishbone (high mount). The descriptions of each type and geometrical condition are as follows:


\begin{figure}[htb]
\centering
 \includegraphics[width=0.5\textwidth]{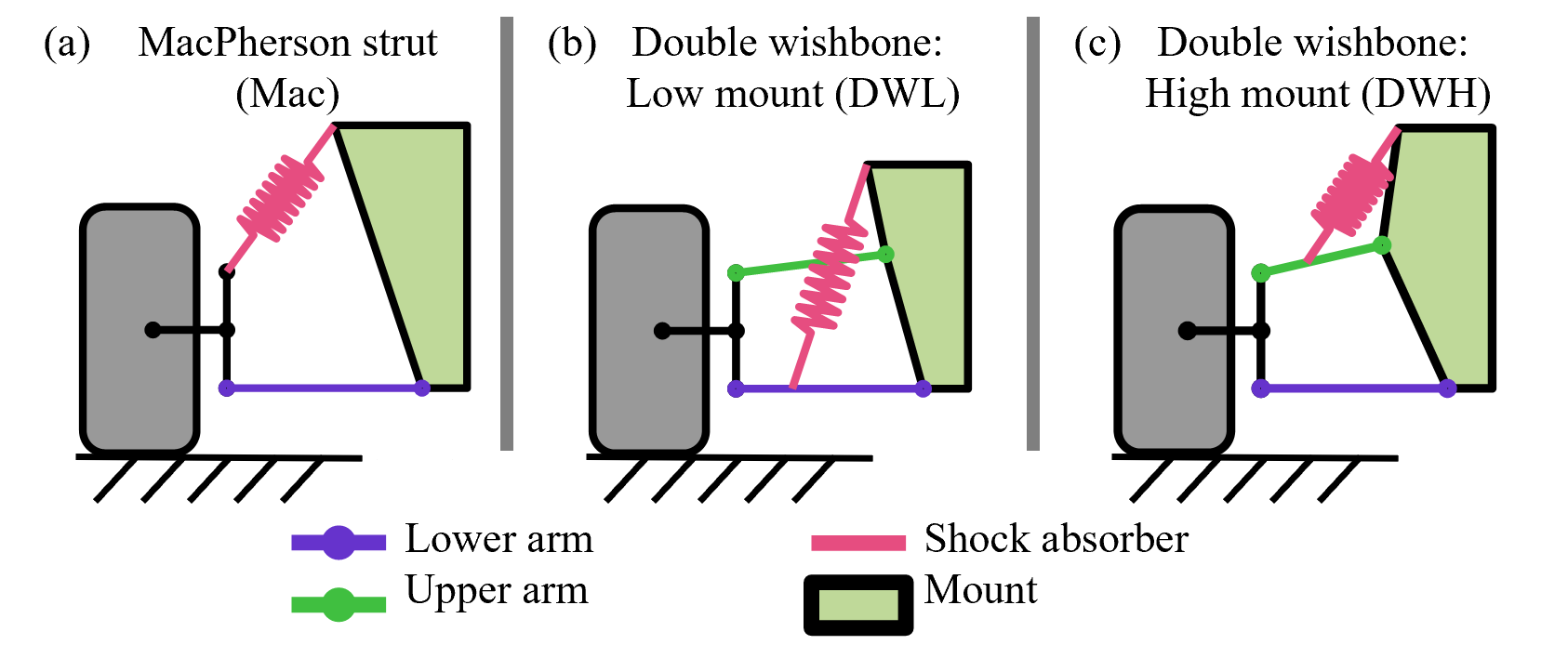}
\caption{Suspension mechanism type definition}
\label{fig:mechanism type definition}
\vspace{-9pt}
\end{figure}

\subsubsection{Type 1: MacPherson strut suspension}\label{subsec3_1_1}
The MacPherson strut(in Fig. \ref{fig:mechanism type definition} (a)) is one of the simplest forms of suspension systems, commonly used for front wheels. Its design is advantageous due to the reduced number of components and a relatively straightforward structure, making it easy to manufacture and cost-effective. The MacPherson strut comprises a single strut (integrating the spring and shock absorber) and a lower control arm, offering benefits in space savings and weight reduction. However, it provides relatively low lateral rigidity, making it less suitable for high-performance vehicles.

\begin{figure}[htb]
\centering
 \includegraphics[width=0.5\textwidth]{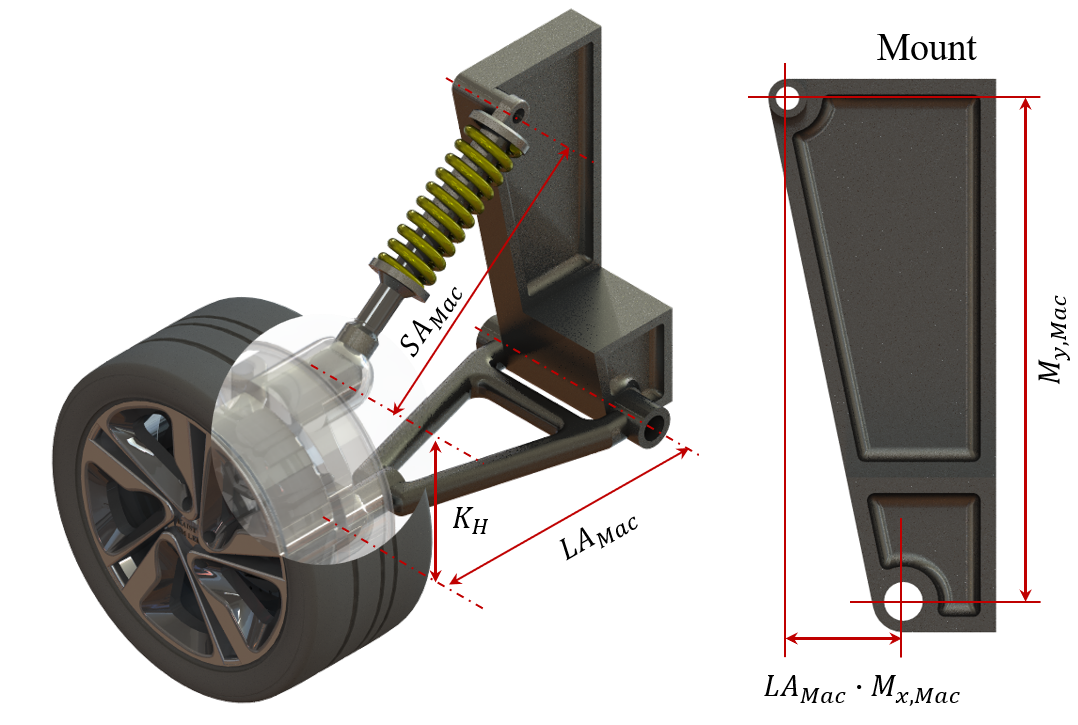}
\caption{MacPherson strut suspension design parameter and 3D configuration}
\label{fig:Mac_parameter}
\vspace{-9pt}
\end{figure}

For the MacPherson strut type, four design parameters are shown in Fig. \ref{fig:Mac_parameter}. Three of them are independent variables, and one is a dependent variable based on the calculation of the independent variables. $LA_{Mac}$ is the length of the lower arm. $M_{x, Mac}$, which has a value between 0 and 1 to prevent infeasible mechanism generation, calculates the $x$ position where the shock absorber is fixed to the mount by multiplying $LA_{Mac}$. $M_{y, Mac}$ is the mounting height between the shock absorber and the lower arm. $SA_{Mac}$, the length of the shock absorber, is derived from the geometric relationship between $LA_{Mac}$, $M_{y, Mac}$, and $M_{x, Mac}$, making it a dependent variable calculated from the other parameters. The equation for calculating the length of $SA_{Mac}$ is provided in Eq.~\eqref{eq_SA_Mac}.

\begin{equation}
SA_{Mac} = \sqrt{(LA_{Mac}(1-M_{x, mac}))^2 + (M_{y, mac}-K_{H})^2} 
\label{eq_SA_Mac}
\end{equation}
, where $K_H$ is the constant knuckle height.

\subsubsection{Type 2: Double wishbone suspension (low mount)}\label{subsec3_1_2}
The double wishbone suspension(in Fig. \ref{fig:mechanism type definition} (b)) consists of two A-shaped control arms (upper and lower arms), allowing for more precise control of wheel movement. In the low mount type, the shock absorber is attached to the lower arm, providing a lower center of gravity and stable steering characteristics. Despite the complexity and higher manufacturing cost, the double wishbone suspension offers superior handling and ride comfort, making it ideal for high-performance vehicles.

\begin{figure}[h!]
\centering
 \includegraphics[width=0.5\textwidth]{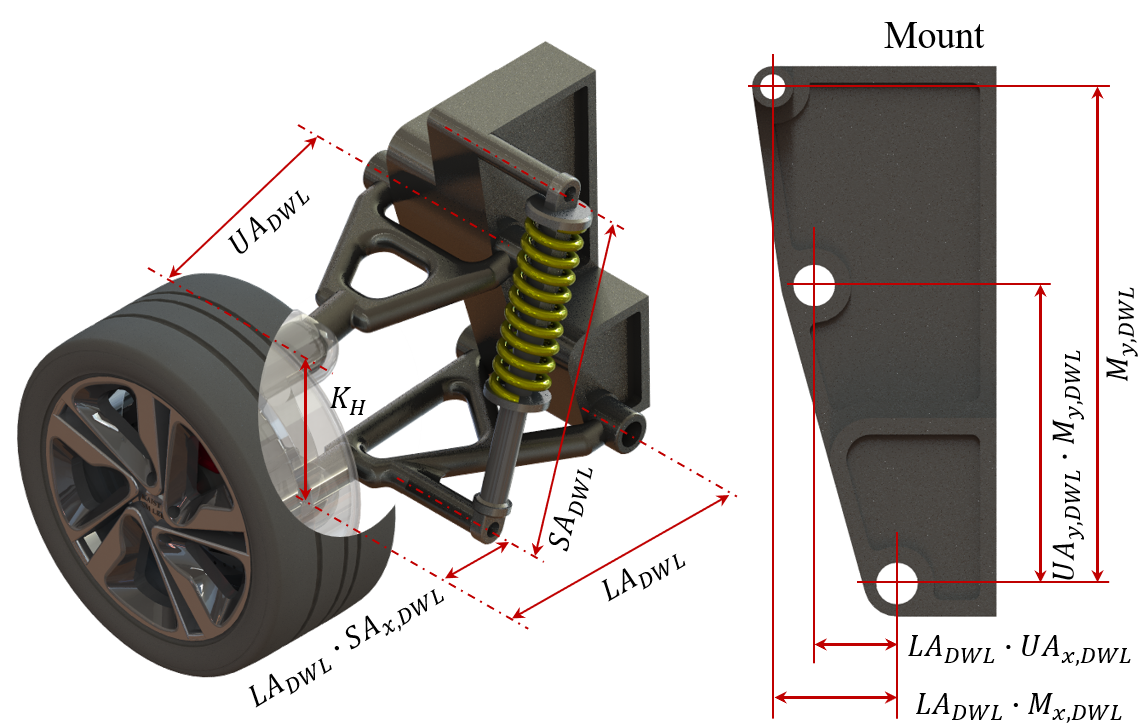}
\caption{Double Wishbone (low mount) suspension design parameter and 3D configuration}
\label{fig:DWL_parameter}
\vspace{-9pt}
\end{figure}

For the Double wishbone (low mount) type, there are eight design parameters(in Fig. \ref{fig:DWL_parameter}): six independent variables and two dependent variables. $LA_{DWL}$ is the length of the lower arm. $SA_{x, DWL}$, a ratio value, calculates the $x$ position where the shock absorber is fixed by multiplying it with $LA_{DWL}$. $UA_{y, DWL}$, a ratio value to calculate the $x$ position where the upper arm is fixed to the mount, calculated by multiplying $LA_{DWL}$. $UA_{y, DWL}$ is the ratio value to calculate the $y$ position where the upper arm is fixed to the mount, calculated by multiplying $M_{y, DWL}$. $M_{x, DWL}$ is the ratio for calculating the $x$ position where the shock absorber is fixed to the mount, calculated by multiplying the ratio with $LA_{DWL}$. $M_{y, DWL}$ represents the total height of the mount.

$UA_{DWL}$ and $SA_{DWL}$, the lengths of the upper arm and the shock absorber respectively, are dependent variables derived from the geometric relationships between the six independent variables. The equations for determining the lengths of $UA_{DWL}$ and $SA_{DWL}$ are provided in Eq.~\eqref{eq_UA_DWL} and ~\eqref{eq_SA_DWL}.

\begin{equation}
UA_{DWL} = \sqrt{(LA_{DWL}(1-UA_{x, DWL}))^2 + (M_{y, DWL} \cdot UA_{x, DWL})^2}
\label{eq_UA_DWL}
\end{equation}

\begin{equation}
SA_{DWL} = \sqrt{(LA_{DWL}(1-SA_{x, DWL}-M_{x, DWL})^2 + (M_{y, DWL}-K_{H})^2} 
\label{eq_SA_DWL}
\end{equation}

\subsubsection{Type 3: Double wishbone suspension (high mount)}\label{subsec3_1_3}
The high mount type of double wishbone suspension(in Fig. \ref{fig:mechanism type definition}(c)) features the shock absorber attached to the upper arm. This configuration extends the range of motion for the suspension and enhances performance across various driving conditions. High mount suspensions are predominantly used in racing cars or high-performance sports cars, where maximizing stability during high-speed driving is crucial.

\begin{figure}[htb]
\centering
 \includegraphics[width=0.5\textwidth]{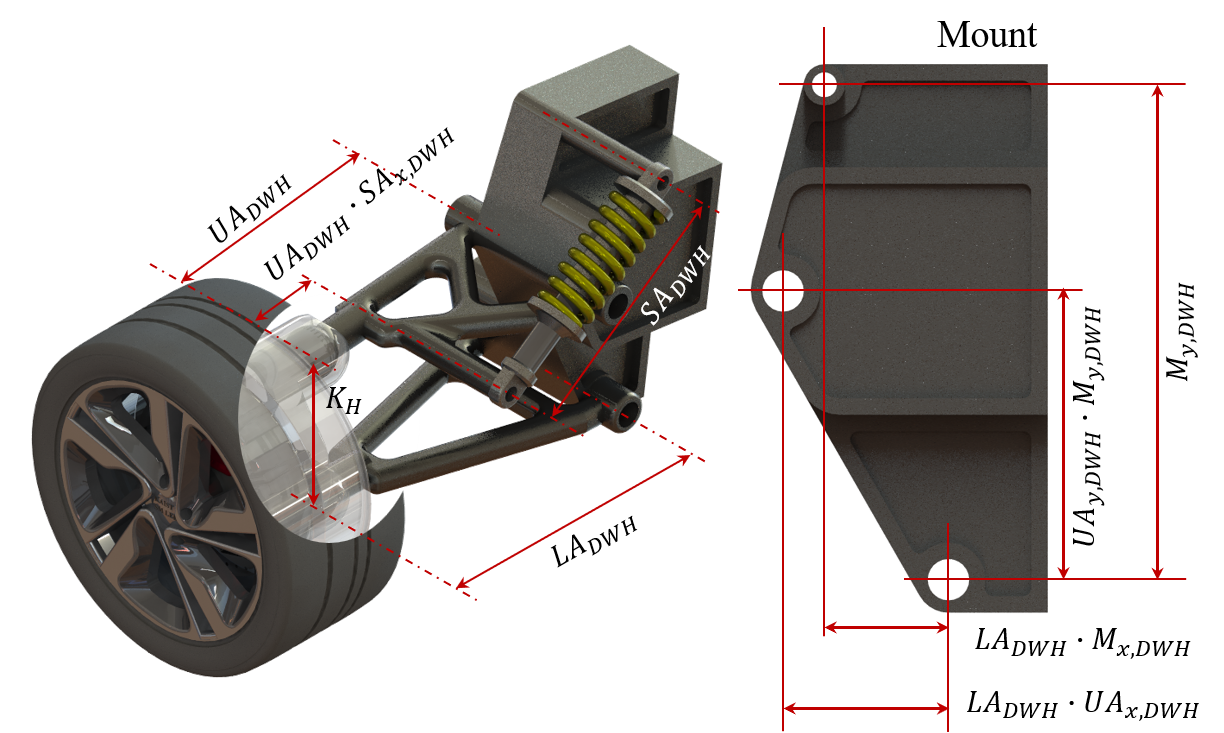}
\caption{Double wishbone (high mount) suspension design parameter and 3D configuration}
\label{fig:DWH_parameter}
\vspace{-9pt}
\end{figure}

For the Double wishbone (high mount) type, there are eight design parameters(in Fig. \ref{fig:DWH_parameter}): six independent variables and two dependent variables. $LA_{DWH}$ is the length of the lower arm. $SA_{x, DWH}$, a ratio value, calculates the $x$ position where the shock absorber is fixed by multiplying it with $UA_{DWL}$. $UA_{x, DWH}$, a ratio value to calculate the $x$ position where the upper arm is fixed to the mount, calculated by multiplying $LA_{DWH}$. $UA_{y, DWH}$ is the ratio value to calculate the $y$ position where the upper arm is fixed to the mount, calculated by multiplying $M_{y, DWH}$. $M_{x, DWH}$ is the ratio for calculating the $x$ position where the shock absorber is fixed to the mount, calculated by multiplying the ratio with $LA_{DWH}$. $M_{y, DWH}$ represents the total height of the mount.

$UA_{DWL}$ and $SA_{DWL}$, the lengths of the upper arm and the shock absorber respectively, are dependent variables derived from the geometric relationships between the six independent variables. The equations for determining the lengths of $UA_{DWL}$ and $SA_{DWL}$ are provided in Eq.~\eqref{eq_UA_DWH} to ~\eqref{eq_SA_DWH}.

\begin{equation}
UA_{DWH} = \sqrt{(LA_{DWH}(1-UA_{x, DWH}))^2 + (M_{y, DWH} \cdot UA_{x, DWH})^2} 
\label{eq_UA_DWH}
\end{equation}

To calculate the length of the shock absorber, it is first necessary to calculate the $x$ and $y$ positions that have been fixed to the upper arm. These positions can be calculated by combining the declared design parameters, and the process is as follows:

\begin{align}
x_{1,SA,DWH} &= LA_{DWH}\cdot SA_{x, DWH}(1-UA_{x, DWH}) \\
y_{1,SA,DWH} &= SA_{x, DWH}\cdot(M_{y, DWH} \cdot UA_{y, DWH} - 2 \cdot K_{H}) + K_{H}\\
x_{2,SA,DWH} &= LA_{DWH}(1-M_{x, DWH}) \\
y_{2,SA,DWH} &= M_{y, DWH} - K_{H}\\
SA_{DWH} &= \sqrt{(x_{2,SA,DWH}- x_{1,SA,DWH})^2 + (y_{2,SA,DWH}-y_{1,SA,DWH})^2}
\label{eq_SA_DWH}
\end{align}

\subsection{Stage 2: Automation of parametric design generation}\label{subsec3_2}
To generate design variables for each suspension type, the Latin Hypercube Sampling (LHS) technique was employed. LHS is a statistical method that evenly samples the design space, ensuring that the possible values of each variable are uniformly distributed. This approach allows for a wide range of design variable combinations, facilitating the exploration of numerous design possibilities. Once the design variables were generated using LHS, these variables were utilized to automatically create 3D CAD models. For this purpose, Rhinoceros 3D® software was used (\citeauthor{mcneel-rhinoceros}). Rhino Grasshopper is a powerful parametric design tool that enables the automatic generation of 3D models based on input variables. As mentioned in Stage 1, each type of suspension design parameter description and sampling range is shown in Table~\ref{table_LHS}.

\begin{table}[h]
\caption{Each type suspension design parameter description and LHS range}
\label{table_LHS}
\begin{tabular*}{\textwidth}
{ p{0.05\linewidth}  p{0.08\linewidth} p{0.06\linewidth}  p{0.53\linewidth}   p{0.05\linewidth}  p{0.06\linewidth}}
\toprule%
\multirow{2}{=}{Type} & \multirow{2}{=}{Design variable} &  \multirow{2}{*}{Unit} & \multirow{2}{*}{Description} & \multicolumn{2}{@{}c@{}}{Design range} \\\cmidrule{5-6}%
  &     &  & & min & max  \\
\Xhline{2pt}
\multirow{6}{=}{\centering  $\textbf{x}_{Mac}$ } 
& $LA_{Mac}$    & [mm] & Lower arm length & 350 & 700 \\
& $M_{x, Mac}$  & ratio & Ratio value, to calculate the $x$ position where the shock absorber is fixed to the mount by multiplying $LA_{Mac}$  & 0.0  & 0.8 \\
& $M_{y, Mac}$  & [mm] & $y$ position where the mount and the shock absorber are fixed & 800 & 1000 \\\cmidrule{2-6}
& $SA_{Mac}$   & [mm] & The length of the shock absorber  & \multicolumn{2}{@{}c@{}}{dependent variable} \\
\Xhline{2pt}

\multirow{14}{=}{\centering  $\textbf{x}_{DWL}$ } 
& $LA_{DWL}$     & [mm] & Lower arm length & 350 & 700 \\
& $SA_{x, DWL}$  & ratio & Ratio value, to calculate the $x$ position where the shock absorber is fixed by multiplying $LA_{DWL}$ & 0.28 & 0.8 \\
& $UA_{x, DWL}$  & ratio & Ratio value, to calculate the $x$ position where the upper arm is fixed to the mount by multiplying $LA_{DWL}$ & 0.1 & 0.7 \\
& $UA_{y, DWL}$  & ratio & Ratio value, to calculate the $y$ position where the upper arm is fixed to the mount multiplying $M_{y, DWL}$& 0.6 & 0.9 \\
& $M_{x, DWL}$   & ratio & Ratio value, to calculate the $x$ position where the shock absorber is fixed to the mount by multiplying $LA_{DWL}$ & 0.14 & 0.8 \\
& $M_{y, DWL}$   & [mm] & $y$ position where the mount and the shock absorber are fixed & 500 & 1000 
\\\cmidrule{2-6}
& $UA_{DWL}$ &  [mm] & Upper arm length  & \multicolumn{2}{@{}c@{}}{dependent variable} \\
& $SA_{DWL}$ &  [mm] & Shock absorber length  & \multicolumn{2}{@{}c@{}}{dependent variable} \\
\Xhline{2pt}

\multirow{14}{=}{\centering  $\textbf{x}_{DWH}$ } 
& $LA_{DWH}$     & [mm] & Lower arm length & 350 & 700 \\
& $SA_{x, DWH}$  & ratio & Ratio value, to calculate the $x$ position where the shock absorber is fixed by multiplying $UA_{DWH}$ & 0.28 & 0.8 \\
& $UA_{x, DWH}$  & ratio & Ratio value, to calculate the $x$ position where the upper arm is fixed to the mount by multiplying $LA_{DWH}$  & 0.1 & 0.7 \\
& $UA_{y, DWH}$  & ratio & Ratio value, to calculate the $y$ position where the upper arm is fixed to the mount multiplying $M_{y, DWH}$& 0.6 & 0.9 \\
& $M_{x, DWH}$   & ratio & Ratio value, to calculate the $x$ position where the shock absorber is fixed to the mount by multiplying $LA_{DWH}$ & 0.14 & 0.8 \\
& $M_{y, DWH}$   & [mm] & $y$ position where the mount and the shock absorber are fixed & 500 & 1000 
\\\cmidrule{2-6}
& $UA_{DWH}$   & [mm] & Upper arm length  & \multicolumn{2}{@{}c@{}}{dependent variable} \\
& $SA_{DWH}$   & [mm] & Shock absorber length  & \multicolumn{2}{@{}c@{}}{dependent variable} \\
\Xhline{2pt}
\botrule
\end{tabular*}
\footnotetext{Note: To prevent the generation of infeasible cases when generating mechanisms, some design parameters are defined as the ratio to the length and implemented utilizing multiplication, whereby the length variable is applied. Dependent design variables are not subjected to LHS and are not directly input into model training or the optimization problem. This is because they can be derived by calculating the geometric relationships among other independent variables. These calculations are necessary for parametric design automation and assembling the bodies in the dynamic analysis automation.}
\end{table}

In practical research scenarios, it is often not feasible to utilize the entirety of available computational resources or sample spaces, and hence, a strategic reduction in sample size is necessary. In this study, although the optimal conditions suggest an extensive dataset for solving the optimization problems effectively, we pragmatically constrained our samples to a manageable number for computational efficiency and practical feasibility. Specifically, for the suspension system problem at hand, we have deliberately chosen to limit the number of samples to ensure a balance between computational efficiency and the quality of the optimization results. For the MacPherson strut suspension type, 300 samples were generated; for the double wishbone low mount type, 209 samples were used; and for the double wishbone high mount type, 200 samples were selected. This strategic selection of sample sizes was made to maintain a reasonable computational burden while still providing a robust exploration of the design space. By choosing these specific sample sizes, we ensure that the optimization process remains efficient and manageable, reflecting a pragmatic approach in real-world applications.

\subsection{Stage 3: Dynamic analysis automation}\label{subsec3_3}
In Stage 3, the objective is to analyze the dynamic behavior of each suspension system, which has been converted into a 3D CAD model. The data obtained from these analyses will be employed as labels for surrogate model training in Stage 4.

The suspensions generated in Stage 1 and 2 were randomly selected within a limited LHS range, consequently lacking insight into their physical performance. For relatively simple mechanisms like the MacPherson strut suspension, an exact solution can be derived using the governing equations of a 2-mass system. However, in the case of actual suspensions, the assumptions required to formulate differential equations are more complex. Additionally, for double wishbone suspensions, the nonlinear displacements resulting from the mechanical structure suggest that solving the problem using governing equations is inherently limited. Therefore, to achieve an optimal mechanism that considers performance, design optimization through a surrogate model is a more effective approach.

\subsubsection{Dynamic analysis setup}\label{subsec3_3_1}
\begin{figure}[htb]
\centering
 \includegraphics[width=0.5\textwidth]{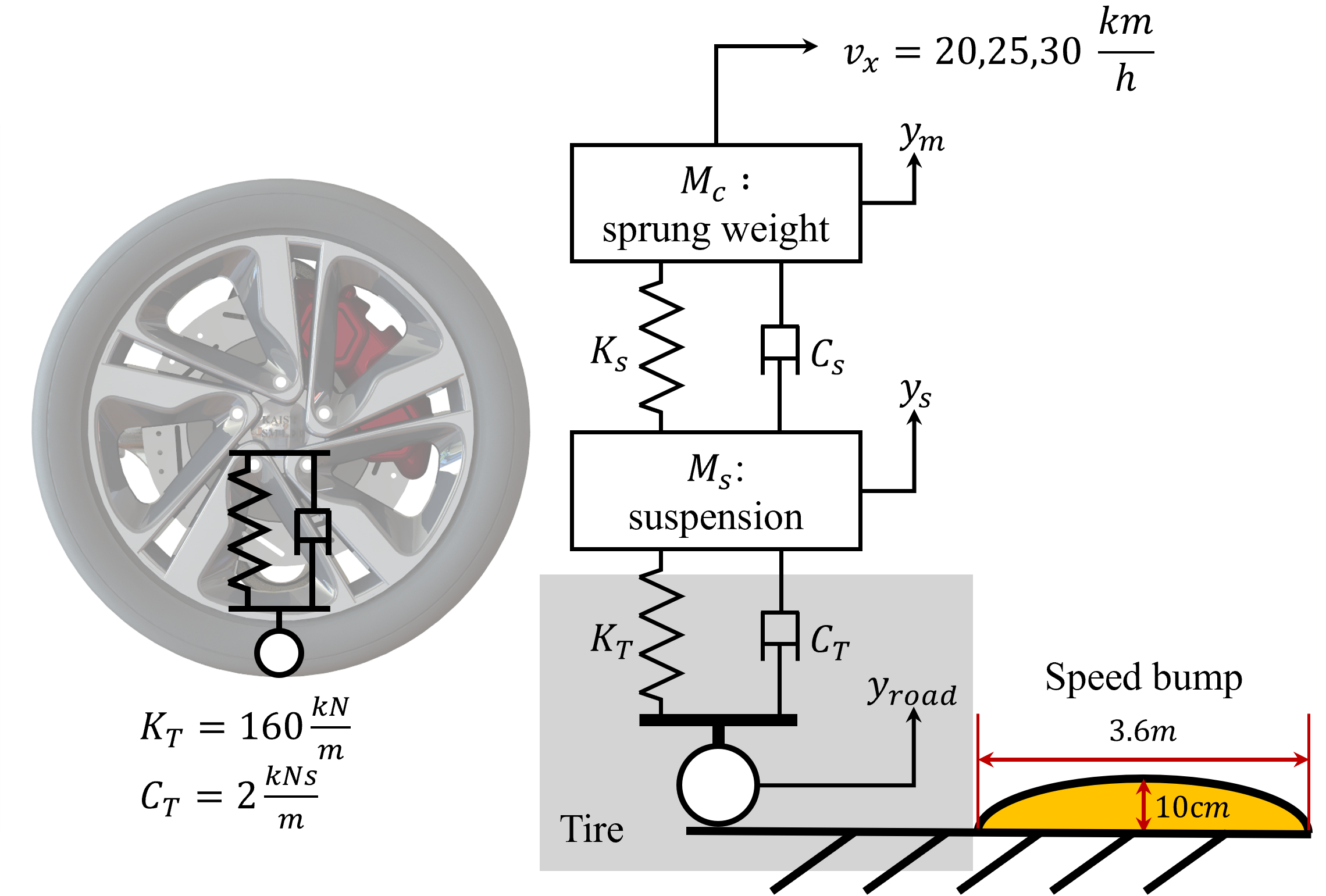}
\caption{Tire and road condition configuration}
\label{fig:Dynamic_setting_diagram}
\vspace{-9pt}
\end{figure}

In the dynamic analysis, the mount sprung mass ($M_c$) and the shock absorber's damping coefficient ($C_s$) and stiffness ($K_s$) were varied. The MCK parameters mentioned above were then input into the dynamic simulation software RecurDyn 2023. 50 combinations of the selected parameters were generated using the LHS method (Table~\ref{table_mck_LHS}). Accordingly, for each generated suspension mechanism, dynamic analyses were conducted by varying the values of the shock absorber and mount by the 50 conditions mentioned above. The tire stiffness ($K_T$) and damping coefficients ($C_T$) were also specified for the dynamic analysis, with typical values ranging from 150 to 300$kN/m$ for stiffness and 1.5 to 3$kNs/m$ for damping (\citeauthor{pacejka2005tire}). The proposed solver setting used values of $K_T = 160kN/m$ for stiffness and $C_T = 2kNs/m$ for damping(in Fig.~\ref{fig:Dynamic_setting_diagram}).

\begin{table}[h]
\caption{MCK parameter design range (LHS)}
\label{table_mck_LHS}%
\begin{tabular}{l c c c }
\toprule
Parameter & $M_c$  & $C_s$ & $K_s$\\
Unit & [kg]  & [$\frac{kN \cdot s}{m}$] & [$\frac{kN}{m}$] \\
\midrule
min   & 300   & 5   & 50  \\
max   & 400   & 20  & 150  \\
\botrule
\end{tabular}
\end{table}


To predict the performance metrics of the sprung mass ($M_c$) and the shock absorber ($C_s \& K_s$) for each suspension model, dynamic analyses were conducted under various conditions. The dynamic analysis assumes the vehicle drives over a typical speed bump at 20, 25, and 30$km/h$. The speed bump used in the analysis has a convex shape with a width of 3.6$m$ and a 10$cm$ height, representing common dimensions found in real-world environments(in Fig.~\ref{fig:Dynamic_setting_diagram}). This profile was applied to the ground in the dynamic simulations to ensure realistic testing conditions. These conditions were used to simulate real-world driving scenarios and to evaluate the performance of the suspensions under different speeds (\citeauthor{godbole2021vehicle,xiao2020influence}).

\subsubsection{Low-fidelity analysis: Multibody Dynamics(MBD)}\label{subsec3_3_2}
This study distinguishes between low-fidelity and high-fidelity analyses to establish a multi-fidelity approach. In this section, we perform a low-fidelity analysis using Multibody Dynamics (MBD) to evaluate the suspension systems under various conditions. This analysis uses rigid body dynamics, where each part of the suspension is assumed to be a rigid body. The dynamic response of each suspension mechanism as it travels over a speed bump is analyzed to extract three performance metrics(in Fig.~\ref{fig:MBD_MFBD_analysis} (a)):

Maximum suspension travel ($P_{max}$): This metric measures the maximum relative vertical displacement of the mount, excluding any bias from the initial vehicle height. It is determined by calculating the difference between the maximum and minimum vertical positions of the mount as the vehicle passes over the speed bump. This provides insight into the suspension's ability to absorb shocks and maintain vehicle stability.

Maximum acceleration ($a_{max}$): This metric captures the peak acceleration experienced by the mount. It is crucial for assessing the ride comfort and the impact of sudden shocks on the vehicle's structure and passengers.

First mode natural frequency($\omega_{n}$): In addition to the other metrics, we conduct an eigenvalue analysis to determine the natural frequencies of the suspension systems. Our specific focus is on the first mode of vibration, which provides a deep understanding of the system's inherent dynamic behavior. This frequency serves as a reference for understanding the system's resonance characteristics and potential susceptibility to vibration issues.

By analyzing these three metrics, we aim to assess the suspension dynamic performance comprehensively. The maximum suspension travel and acceleration directly measure the system's response to external disturbances, while the natural frequency offers insight into the system's inherent dynamic behavior. 


\subsubsection{High-fidelity analysis: Multi-Flexible Body Dynamics(MFBD)}\label{subsec3_3_3}
\begin{figure}[h!]
\centering
 \includegraphics[width=1\textwidth]{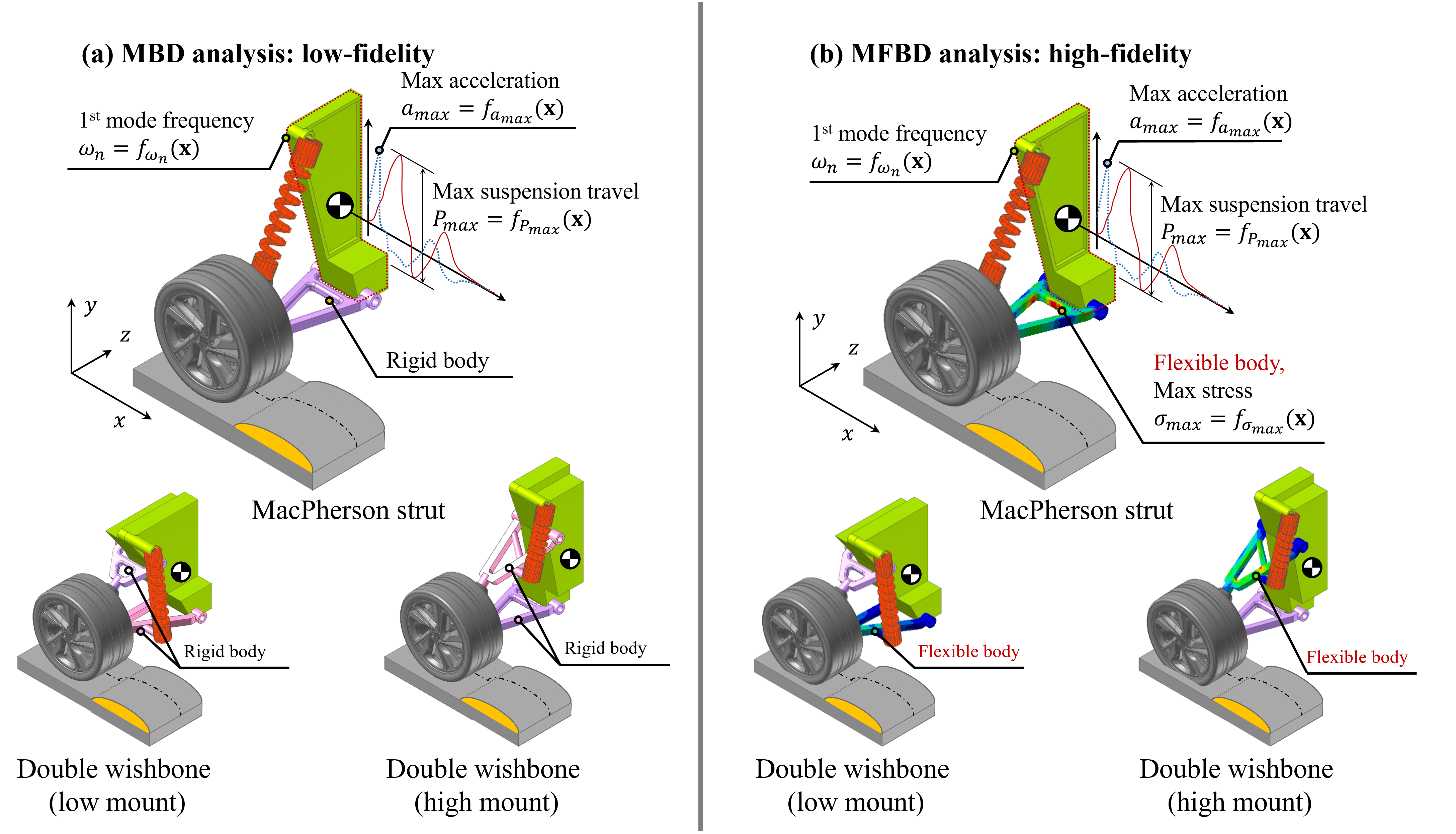}
\caption{MBD and MFBD analysis configuration}
\label{fig:MBD_MFBD_analysis}
\vspace{-9pt}
\end{figure}

In this section, we address the high-fidelity analysis which includes performance that cannot be considered in rigid body analysis. In MBD, each component is assumed to be a rigid body, and thus any deformation within the components is ignored. Therefore, the consideration of the dynamic behavior of mechanisms with flexible bodies is not possible, which means that it cannot realistically reflect the exact results. To evaluate the modifications to dynamic behavior and the stresses that arise in each member when components are regarded as flexible, the Multi-Flexible Body Dynamics (MFBD) analysis is introduced(in Fig.~\ref{fig:MBD_MFBD_analysis} (b)).

The MFBD analysis includes not only the three performance metrics ($P_{max}$, $a_{max}$, $\omega_{n}$) discussed in the MBD but also the maximum stress ($\sigma_{max}$) of the flexible body(finite element). This approach allows for a more detailed assessment of the dynamic responses and stress distribution within the suspension, which is crucial for identifying potential failure points and ensuring the reliability of the design.
The material properties used for the components in the dynamic analysis are based on SAE 9255 steel, which is commonly used in automotive suspension systems. The Young's modulus($E_{a}$) for this steel is $210 GPa$.



\subsection{Stage 4: Multi-fidelity surrogate model}\label{subsec3_4}
\begin{figure}[htb]
\centering
 \includegraphics[width=1\textwidth]{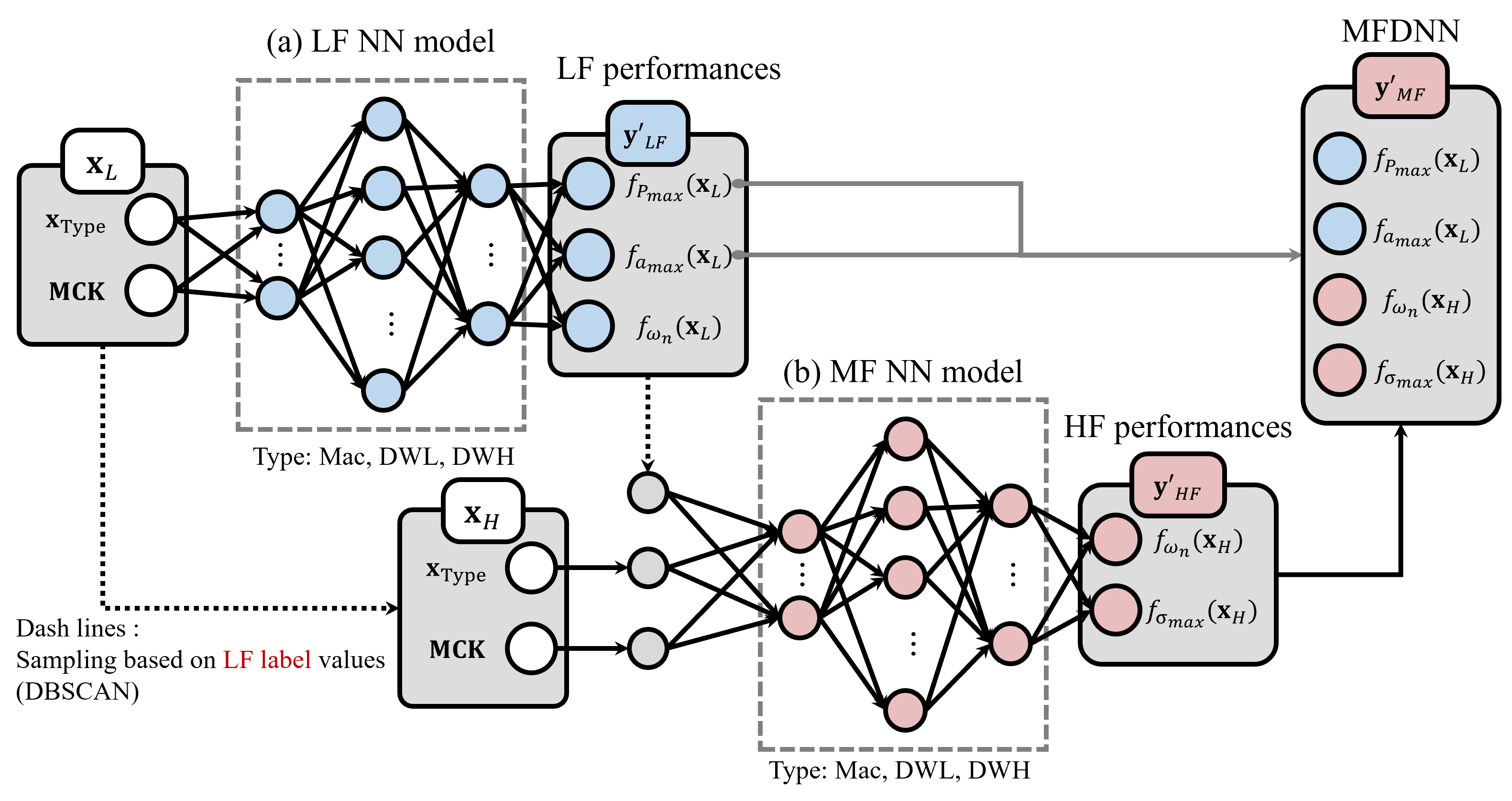}
\caption{MF model diagram}
\label{fig:Multi_fidelity_diagram}
\vspace{-9pt}
\end{figure}

Stage 4 focuses on training a multi-fidelity surrogate model to efficiently predict the dynamic performance metrics($P_{max}$, $a_{max}$, $\omega_{n}$) of each type of suspension mechanism and the maximum stress($\sigma_{max}$) of flexible bodies(in Fig.~\ref{fig:Multi_fidelity_diagram}). However, since the defined four performances exhibit non-linearity due to the geometric complexity of each suspension type, we opt for the multi-layer perceptron (MLP) approach. The ability of MLP to capture the nonlinear relationships between design variables and performance metrics provides significant advantages in the engineering domain.

\subsubsection{Low-fidelity model}\label{subsec3_4_1}
The low-fidelity model focuses on predicting three performance metrics using the independent design variables ($\textbf{x}_{Mac},\textbf{x}_{DWL},\textbf{x}_{DWH}$) and MCK parameters as inputs(in Fig.~\ref{fig:Multi_fidelity_diagram} (a)). These metrics include $P_{max}$, $a_{max}$, and $\omega_{n}$. Utilizing the low-fidelity analysis results obtained in Stage 3, the low-fidelity model provides a first approximation of the suspension performance metrics.

To effectively integrate the low fidelity analysis (MBD) with the high-fidelity analysis (MFBD), the density-based spatial clustering of applications with noise (DBSCAN) algorithm was utilized(\citeauthor{yi2020multi}). This approach enabled the extraction of a representative subset of samples from the MBD results. Specifically, a total of 5$\%$ of the samples from the set $\textbf{x}_{L}$, which was utilized in the low-fidelity analysis, were selected for the MFBD. The representative samples, $\textbf{x}_{H}$, was then utilized for the MFBD, ensuring that the most informative and diverse samples were sampled while ensuring computational costs remained within an acceptable range.

\subsubsection{Multi-fidelity model}\label{subsec3_4_2}
The multi-fidelity model enhances the predictive capability by utilizing the outputs of the low-fidelity model as supplementary inputs. In particular, the multi-fidelity model employs the independent design variables ($\textbf{x}_{Mac},\textbf{x}_{DWL},\textbf{x}_{DWH}$), MCK parameters, and the three performance metrics predicted by the low-fidelity model to not only anticipate the aforementioned performance metrics ($P_{max}$, $a_{max}$, and $\omega_{n}$) but also to project the maximum stress value ($\sigma_{max}$) in the flexible body(in Fig.~\ref{fig:Multi_fidelity_diagram} (b)). This approach addresses the limitations of the low-fidelity model by refining the predictions of performance metrics that exhibit discrepancies between low-fidelity and high-fidelity analyses and by providing predictions for performance metrics that cannot be obtained from low-fidelity analysis alone.


\subsection{Stage 5: Design optimization}\label{subsec3_5}

The aim of Stage 5 is to optimize the performance metrics of each suspension type. Due to the need for multiple evaluations during the optimization process, we utilize the multi-fidelity model trained in Stage 4 to predict performance values for these evaluations. The proposed application focuses on four performance metrics: $P_{max}$, $a_{max}$, $\omega_{n}$, and $\sigma_{max}$. However, for vehicle suspension systems, the first natural frequency mode ($\omega_{n}$) is used as a reference performance metric rather than an objective function. This is because the natural frequency is crucial for understanding the dynamic behavior and potential resonance issues of the suspension system. A suspension system with a natural frequency that matches the frequency of road irregularities can lead to resonance, causing excessive vibrations and reduced ride comfort. Therefore, while $\omega_{n}$ is not directly optimized, it is monitored to ensure it falls within an acceptable range that avoids resonance with common driving conditions. This approach helps in designing a suspension system that balances ride comfort, stability, and structural integrity without explicitly optimizing the natural frequency as a performance metric. Thus, the optimization problem is formulated as a multi-objective optimization (MOO) problem with three performance metrics ($P_{max}$, $a_{max}$, and $\sigma_{max}$) as objective functions.

To solve this optimization problem, we use the well-known non-dominated sorting genetic algorithm (NSGA-II) algorithm as the optimizer, using the Python package \textsc{pymoo} (\citeauthor{blank2020pymoo}). The objectives, constraints, and settings for NSGA-II are as follows:


\subsubsection{Design variables}
The independent design variables for each type of suspension mechanism, as defined in Stages 1 and 2 (in Fig.~\ref{fig:mechanism type definition} and Table~\ref{table_LHS}), are as follows: $\textbf{x}_{Mac}$, $\textbf{x}_{DWL}$, and $\textbf{x}_{DWH}$. Additionally, the MCK parameters (in Table~\ref{table_mck_LHS}) are considered in the optimization process. The design variables for each type are defined as combinations of mechanism design variables and MCK parameters: $\textbf{x} = [\textbf{x}_{Mac}, \textbf{MCK}]$, $\textbf{x} = [\textbf{x}_{DWL}, \textbf{MCK}]$, and $\textbf{x} = [\textbf{x}_{DWH}, \textbf{MCK}]$ for the respective suspension types. These combined design variables allow for the simultaneous optimization of both each type of suspension and the shock absorber's mechanical properties. In some cases, specific values could be fixed or design ranges provided to perform optimal design tailored to different performance requirements.

\subsubsection{Objective functions and constraints}
In the optimization process, the objective functions are defined based on the predicted performance metrics $f_{P_{max}}$, $f_{a_{max}}$, and $f_{\sigma_{max}}$. $f_{P_{max}}(\textbf{x})$ represents the predicted maximum suspension travel. This value is a crucial indicator of the suspension's ability to absorb shocks and maintain stability. A lower value of $P_{max}$ suggests better performance in limiting excessive suspension movement, contributing to vehicle stability and passenger comfort. $f_{a_{max}}(\textbf{x})$ denotes the predicted maximum acceleration experienced by the suspension mount. This metric is essential for evaluating ride comfort and the structural impact on the vehicle. Minimizing $a_{max}$ is desirable as it indicates smoother ride quality and reduced peak forces transmitted to the vehicle's structure. $f_{\sigma_{max}}(\textbf{x})$ is the predicted maximum stress within the flexible body components of the suspension. This performance metric is critical for ensuring the durability and reliability of the suspension system. Lower values of $\sigma_{max}$ indicate that the suspension components can withstand operating loads without experiencing high stress, thereby enhancing longevity and safety. Thus, $f_{P_{max}}$, $f_{a_{max}}$, and $f_{\sigma_{max}}$ are defined as objective functions to minimize.

The number of constraints for optimizing each type of suspension using the multi-fidelity surrogate model is as follows: $g_{Mac,l}:(l=1, \dots, 5)$, $g_{DWL,m}:(m=1, \dots, 5)$, and $g_{DWH,n}:(n=1, \dots, 5)$. Although the design parameters for each suspension type are not shared, the performance metrics obtained from MBD and MFBD analysis are the same. Therefore, the boundary conditions for the objective function of each optimization are outlined in Eq.\eqref{eq_obj_Mac},\eqref{eq_obj_DWL}, and \eqref{eq_obj_DWH} . To ensure that the optimization does not escape the design variable range used in training the surrogate model, the design variables for optimization are proposed within the range utilized in the LHS. $5\%$ of the designs are filtered out as outliers to bypass the singularity of the $\omega_{n}$. We used a z-value of 1.95, assuming a normal distribution. Finally, to recommend the optimized design variables for each type based on the input sprung mass ($M_{t}$), a boundary condition is applied to restrict the range of the $M_{c}$  within the MCK conditions. Therefore, each type of suspension optimization problem can be formulated as follows: 

The optimization formulation for the Mac type is expressed as:
\begin{flalign}
\displaystyle\min_{\textbf{x}} &[f_{P_{max}}(\textbf{x}), f_{a_{max}}(\textbf{x}), f_{\sigma_{max}}(\textbf{x})] \label{eq_obj_Mac} \\
\text{w.r.t. } \textbf{x} &= [\textbf{x}_{Mac}, \textbf{MCK}], \nonumber \\
\textbf{x}_{Mac} &= [ LA_{Mac}, M_{x, Mac}, M_{y, Mac} ], \nonumber \\
\textbf{MCK} &= [ M_c, C_s, K_s ], \nonumber \\
\text{s.t. } &  M_{t} - \epsilon \le M_c \le M_{t} + \epsilon. \nonumber \\
g_{Mac,1}: &\quad f_{P_{max}}(\textbf{x}) \ge 0, \nonumber \\
g_{Mac,2}: &\quad f_{a_{max}}(\textbf{x}) \ge 0, \nonumber \\
g_{Mac,3}: &\quad f_{\sigma_{max}}(\textbf{x}) \ge 0, \nonumber \\
g_{Mac,4}: &\quad f_{\omega_{n}}(\textbf{x}) \ge \text{mean}[\omega_{n}] - z \cdot \text{std}[f_{\omega_{n}}(\textbf{x})], \nonumber \\
g_{Mac,5}: &\quad f_{\omega_{n}}(\textbf{x}) \le \text{mean}[\omega_{n}] - z \cdot \text{std}[f_{\omega_{n}}(\textbf{x})]. \nonumber
\end{flalign}
, where $\epsilon$ is a threshold value used to create a boundary condition by adding to or subtracting from $M_{t}$. In the optimization problem handled in this framework, the range of the sprung mass $M_{c}$ is set narrowly to focus on exploring the optimal solutions for other design variables.

The optimization formulation for the DWL type is expressed as:
\begin{flalign}
\displaystyle\min_{\textbf{x}} &[f_{P_{max}}(\textbf{x}), f_{a_{max}}(\textbf{x}), f_{\sigma_{max}}(\textbf{x})], \label{eq_obj_DWL} \\
\text{w.r.t. } \textbf{x} &= [\textbf{x}_{DWL}, \textbf{MCK}], \nonumber \\
\textbf{x}_{DWL} &= [ LA_{DWL}, SA_{x, DWL}, UA_{x, DWL}, UA_{y, DWL}, M_{x, DWL}, M_{y, DWL} ], \nonumber \\
\textbf{MCK} &= [ M_c, C_s, K_s ], \nonumber \\
\text{s.t. } & M_{t} - \epsilon \le M_c \le M_{t} + \epsilon. \nonumber \\
g_{DWL,1}: &\quad f_{P_{max}}(\textbf{x}) \ge 0, \nonumber \\
g_{DWL,2}: &\quad f_{a_{max}}(\textbf{x}) \ge 0, \nonumber \\
g_{DWL,3}: &\quad f_{\sigma_{max}}(\textbf{x}) \ge 0, \nonumber \\
g_{DWL,4}: &\quad f_{\omega_{n}}(\textbf{x}) \ge \text{mean}[\omega_{n}] -z \cdot \text{std}[f_{\omega_{n}}(\textbf{x})], \nonumber \\
g_{DWL,5}: &\quad f_{\omega_{n}}(\textbf{x}) \le \text{mean}[\omega_{n}] -z \cdot \text{std}[f_{\omega_{n}}(\textbf{x})]. \nonumber 
\end{flalign}

Finally, The optimization formulation for the DWH type is expressed as:
\begin{align}
\displaystyle\min_{\textbf{x}} &[f_{P_{max}}(\textbf{x}), f_{a_{max}}(\textbf{x}), f_{\sigma_{max}}(\textbf{x})], \label{eq_obj_DWH} \\
\text{w.r.t. } \textbf{x} &= [\textbf{x}_{DWH}, \textbf{MCK}], \nonumber \\
\textbf{x}_{DWH} &= [ LA_{DWH}, SA_{x, DWH}, UA_{x, DWH}, UA_{y, DWH}, M_{x, DWH}, M_{y, DWH} ], \nonumber \\
\textbf{MCK} &= [ M_c, C_s, K_s ], \nonumber \\
\text{s.t. } & M_{t} - \epsilon \le M_c \le M_{t} + \epsilon. \nonumber \\
g_{DWH,1}: &\quad f_{P_{max}}(\textbf{x}) \ge 0, \nonumber \\
g_{DWH,2}: &\quad f_{a_{max}}(\textbf{x}) \ge 0, \nonumber \\
g_{DWH,3}: &\quad f_{\sigma_{max}}(\textbf{x}) \ge 0, \nonumber \\
g_{DWH,4}: &\quad f_{\omega_{n}}(\textbf{x}) \ge \text{mean}[\omega_{n}] -z \cdot \text{std}[f_{\omega_{n}}(\textbf{x})], \nonumber \\
g_{DWH,5}: &\quad f_{\omega_{n}}(\textbf{x}) \le \text{mean}[\omega_{n}] -z \cdot \text{std}[f_{\omega_{n}}(\textbf{x})]. \nonumber 
\end{align}
The formulation ensures that the optimization process respects each suspension type's mechanical constraints and objective functions.

\section{Results and discussion}\label{sec4}
\subsection{Dynamic analysis (MBD, MFBD) results}\label{subsec4_1}
In Stages 1 and 2, a total of 35,450 low-fidelity analyses (MBD) were conducted using 300(Mac), 209(DWL), and 200(DWH) samples for each suspension type, respectively, across 50 MCK conditions. The automated dynamic simulation process is summarized as follows: 1) The 3D CAD models generated in Stage 2 are imported. 2) The imported components are automatically assembled according to the joints calculated in Stage 1. 3) MCK conditions and speed bump road references are assigned to the assembled models, followed by the MBD analysis.

High-fidelity analyses (MFBD) samples are required to construct the multi-fidelity surrogate model. Due to the expensive computational cost, using all MBD samples for MFBD analysis is impractical. Typically, $5\%$ to $15\%$ of high-fidelity analyses are performed depending on the application. This study analyzed MBD results to sample the cases for MFBD analysis. The design variables for the MBD analysis were sampled using LHS, as shown in Tables~\ref{table_LHS} and \ref{table_mck_LHS}. However, the performance metrics obtained from MBD are not intuitively evenly distributed across the performance space. Thus, selecting samples for MFBD analysis based on the performance values already obtained from MBD to construct a robust multi-fidelity model is reasonable.

\begin{figure}[h!]
\centering
 \includegraphics[width=1\textwidth]{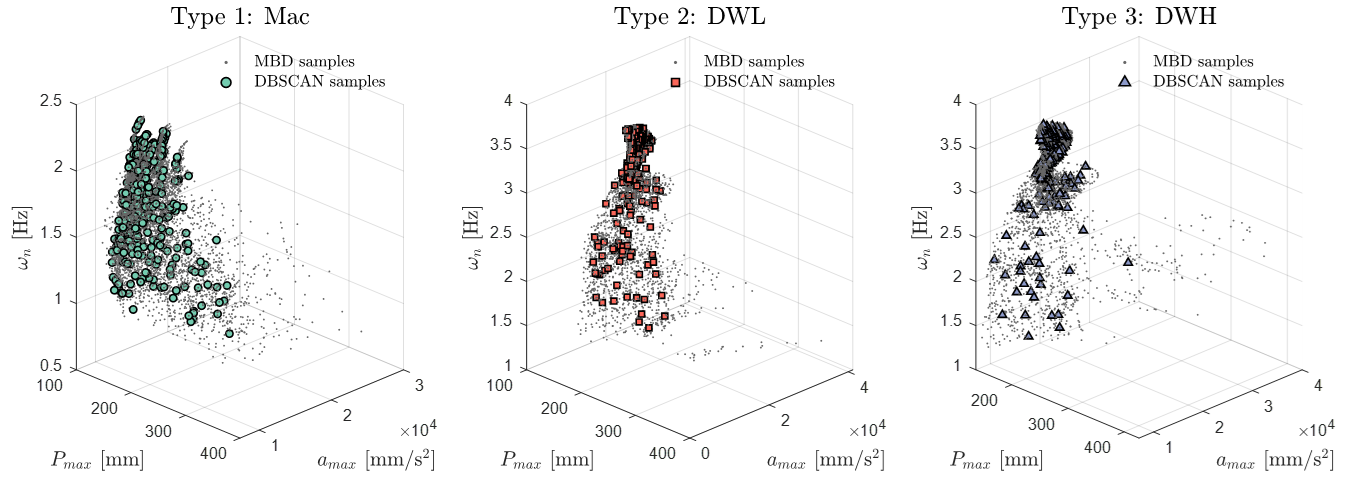}
\caption{MBD samples and DBSCAN samples configuration}
\label{fig:DBSCAN_results}
\vspace{-9pt}
\end{figure}

In this application, the performance metrics obtained from the MBD solver, when plotted in the performance space, displayed both high and low density (Fig.~\ref{fig:DBSCAN_results}). To ensure representative sampling across this non-uniform performance, the DBSCAN method was utilized. Unlike k-means, DBSCAN does not require a predefined number of clusters and automatically groups representative samples based on Euclidean distance. Figure.~\ref{fig:DBSCAN_results} illustrates the $5\%$ reconstructed samples from normalized performance metrics, demonstrating a more evenly distributed sample set.

\begin{figure}[h!]
\centering
 \includegraphics[width=1\textwidth]{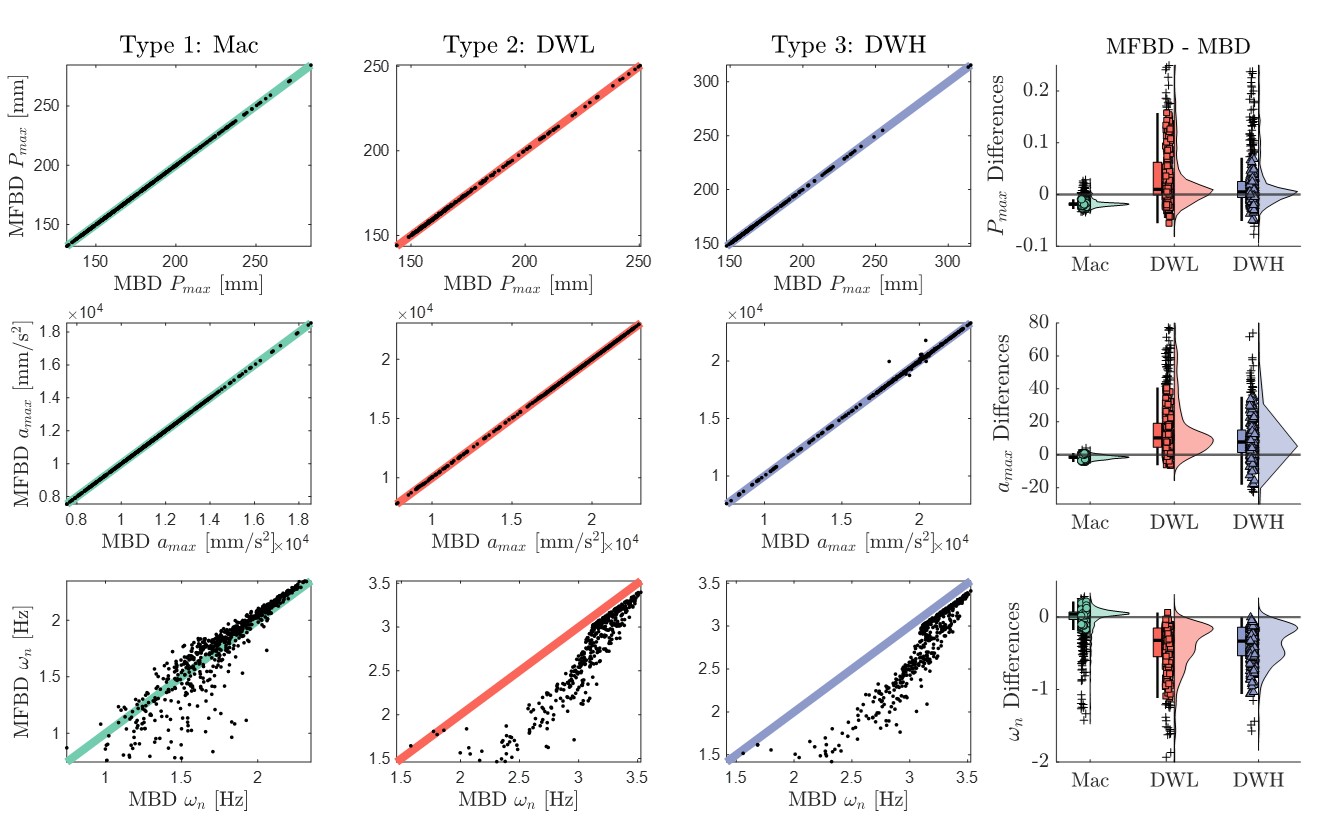}
\caption{MBD vs MFBD results comparison}
\label{fig:MBD_vs_MFBD}
\vspace{-9pt}
\end{figure}

The MFBD analysis of the $5\%$ samples selected via DBSCAN obtained four performance metrics ($P_{max}$, $a_{max}$, $\omega_{n}$, $\sigma_{max}$). Before constructing the multi-fidelity surrogate model, a comparison of three performance metrics ($P_{max}$, $a_{max}$, $\omega_{n}$) between MBD and MFBD analyses is shown in Fig.~\ref{fig:MBD_vs_MFBD}. The $x$-axis represents values from the MBD analysis, and the $y$-axis represents values from the MFBD analysis. The closer the scatter plot is to the $x=y$ line, the smaller the difference between MBD and MFBD analyses. To quantitatively compare the three performance metrics for each type, we calculated the Mean Absolute Percentage Error (MAPE) using Eq.\eqref{eq_MAPE}. MAPE effectively expresses the ratio of multiple data values to errors, making it useful for analyzing trends across various samples, as shown in Fig.~\ref{fig:MBD_vs_MFBD}. The MAPE results for the MBD and MFBD analyses are presented in Table~\ref{tab_MAPE}.

\begin{align}
MAPE &= \frac{100}{N}  \sum_{t=1}^{N} \left| \frac{Y_{MFBD,t} - Y_{MBD,t}}{Y_{MFBD,t}}\right|
\label{eq_MAPE}
\end{align}

\begin{table}[h!]
\caption{MAPE scores each type and performance metrics}
\label{tab_MAPE}
\begin{tabular}{lccc}
\toprule%
     & \multicolumn{3}{@{}c@{}}{MAPE($\%$)} \\\cmidrule{2-4}%
Type &  $P_{max}$ & $a_{max}$ & $\omega_{n}$ \\
\midrule
Mac  & 0.0107 & 0.0146  & \textbf{34.6400} \\
DWL  & 0.0146 & 0.1091  & \textbf{26.9730} \\
DWH  & 0.0227 & 0.1523  & \textbf{16.5128} \\
\botrule
\end{tabular}
\end{table}


The metrics for $P_{max}$ and $a_{max}$ are nearly indistinguishable between the MBD and MFBD analyses, indicating that the results are so closely aligned that additional high-fidelity analyses may be unnecessary for these specific metrics. However, while showing a similar overall trend, the $\omega_{n}$ metric exhibits significant differences in certain cases. This discrepancy is consistent with the finding that the natural frequency of mechanisms assumed to be rigid can be significantly affected when specific parts are modeled as flexible bodies using finite element analysis (\citeauthor{kim2019coupled}).

\subsection{Multi-fidelity model prediction results}\label{subsec4_2}
\begin{figure}[h!]
\centering
 \includegraphics[width=1\textwidth]{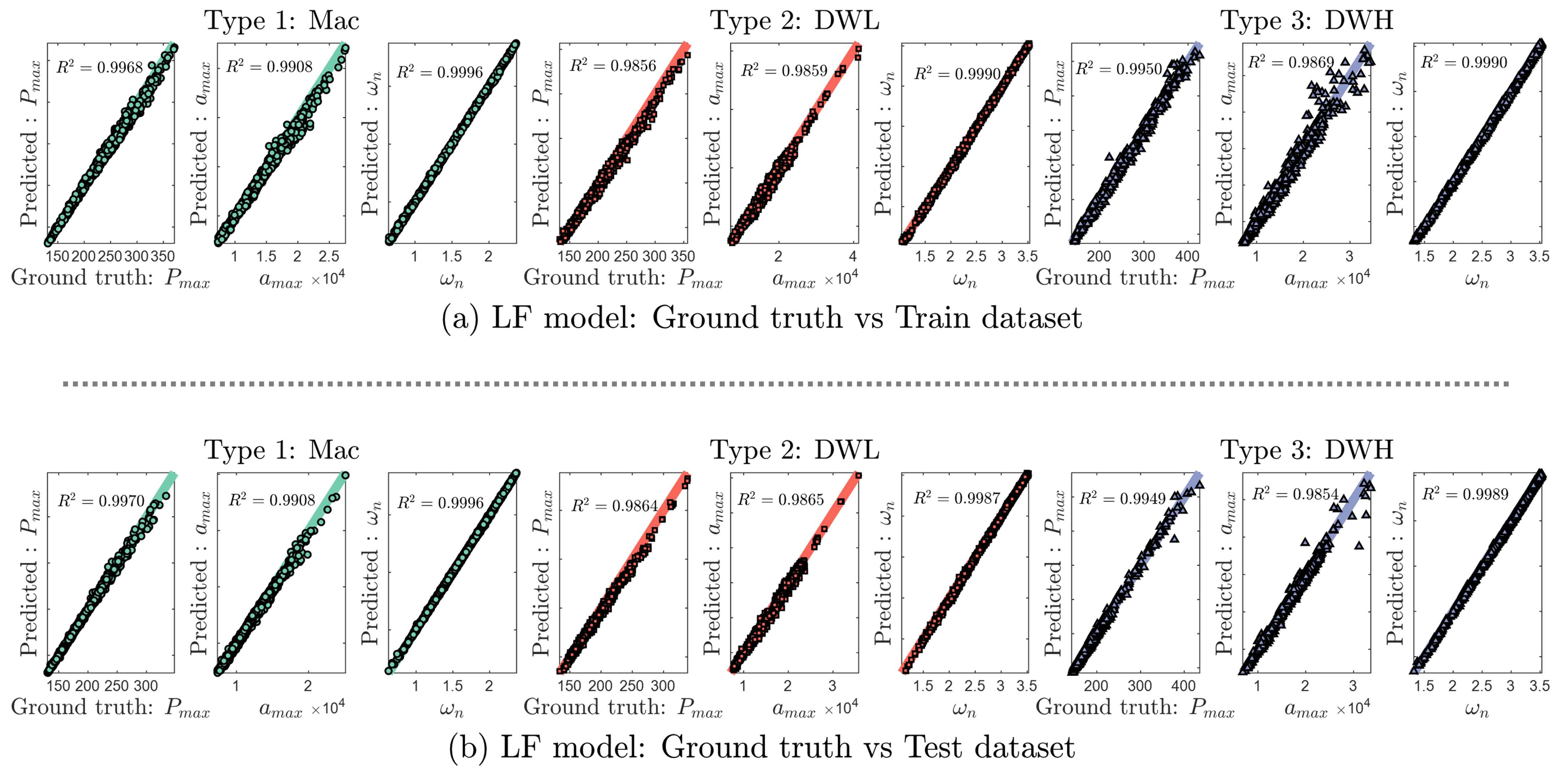}
\caption{Validation of LF model using train(a) and test(b) data: ground truth as x-axis and prediction of LF model as y-axis}
\label{fig:LF_results}
\vspace{-9pt}
\end{figure}

\begin{figure}[h!]
\centering
 \includegraphics[width=1\textwidth]{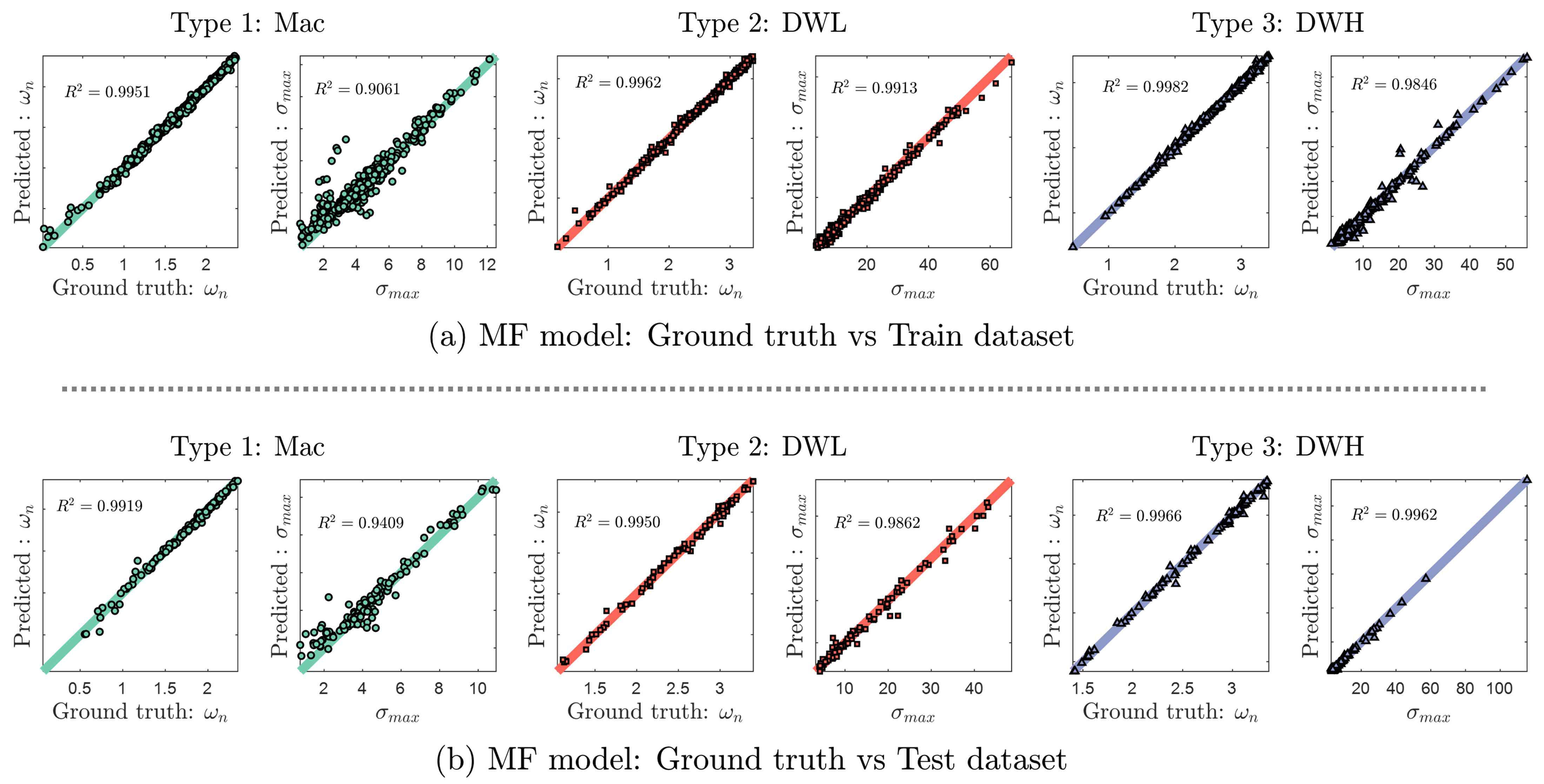}
\caption{Validation of MF model using train(a) and test(b) data: ground truth as x-axis and prediction of MF model as y-axis}
\label{fig:MF_results}
\vspace{-9pt}
\end{figure}

To construct the multi-fidelity (MF) model, it is essential to first train the low-fidelity (LF) model (in Fig.\ref{fig:Multi_fidelity_diagram}). To avoid overfitting during the training of the LF model for each suspension type, an AutoML approach was employed to iterative optimize the hyperparameters. Once the LF model was successfully trained, it served as the foundation for developing the MF model. The MF model was trained using $5\%$ of the samples selected through DBSCAN and the pre-trained LF model. As indicated in Fig.\ref{fig:MBD_vs_MFBD} and Table~\ref{tab_MAPE}, $P_{max}$ and $a_{max}$ show almost identical results between MBD and MFBD analyses; therefore, these metrics were not separately predicted in the MF model. Conversely, due to the notable differences observed in $\omega_{n}$ between the two analyses, it was necessary to predict this metric within the MF model. Additionally, $\sigma_{max}$, a performance metric that cannot be obtained from MBD, was also included in the MF model predictions. The computational environment, hyperparameters used for training the LF and MF models, and the accuracy of the predictions, including the performance of the MF models trained using the $5\%$ samples selected via DBSCAN, are summarized in Table~\ref{tab_Performance of the MLP regressor}, and Fig.~\ref{fig:LF_results},~\ref{fig:MF_results}. The accuracy of the predictions was evaluated using R-squared ($R^2$), Mean Squared Error (MSE), and Root Mean Squared Error (RMSE), providing an objective comparison between the models. Given that most of the trained MLP predictions are similar to the corresponding ground truth values, the MLP model has been trained successfully.



\begin{table}
\caption{Hyperparameter and performance of each MLP regressor}
\label{tab_Performance of the MLP regressor}
\begin{tabular*}{\textwidth}
{ >{\centering\arraybackslash}p{0.04\linewidth} 
  >{\centering\arraybackslash}p{0.05\linewidth} 
  >{\centering\arraybackslash}p{0.04\linewidth} 
  >{\centering\arraybackslash}p{0.17\linewidth} 
  >{\centering\arraybackslash}p{0.05\linewidth} 
  >{\centering\arraybackslash}p{0.05\linewidth} 
  >{\centering\arraybackslash}p{0.05\linewidth} 
  >{\centering\arraybackslash}p{0.09\linewidth} 
  >{\centering\arraybackslash}p{0.09\linewidth} 
  >{\centering\arraybackslash}p{0.09\linewidth}}
\toprule
\multirow{2}{=}{Type} & \multirow{2}{=}{Fidelity type} & \multirow{2}{=}{Batch size} & \multirow{2}{=}{Hidden layer sizes}& \multirow{2}{*}{Epochs} & \multirow{2}{*}{Label} & \multirow{2}{*}{Dataset} & \multirow{2}{*}{$R^2$ score} & \multirow{2}{*}{MSE} & \multirow{2}{*}{RMSE} \\%
  & & & & & & & & & \\
\Xhline{2pt}
\multirow{14}{=}{\centering  Mac } 
& \multirow{8}{=}{\centering  LF (MBD) } & \multirow{8}{=}{\centering  100 } & \multirow{8}{=}{\centering  [128, 64, 32, 16] } & \multirow{8}{=}{\centering  100 }  
        & \multirow{2}{*}{$P_{max}$}    & Train & 0.9968   & $3.28 \cdot 10^1$  & $1.81 \cdot 10^1$ \\
& & & & &                               & Test  & 0.9970   & $2.81 \cdot 10^1$  & $1.68 \cdot 10^1$ \\\cmidrule{6-10}        
& & & & & \multirow{2}{*}{$a_{max}$}    & Train & 0.9908   & $4.15 \cdot 10^4$  & $2.04 \cdot 10^2$ \\
& & & & &                               & Test  & 0.9908   & $3.48 \cdot 10^4$  & $1.87 \cdot 10^2$ \\\cmidrule{6-10} 
& & & & & \multirow{2}{*}{$\omega_{n}$} & Train & 0.9996   & $4.62 \cdot 10^{-5}$  & $6.79 \cdot 10^{-3}$\\
& & & & &                               & Test  & 0.9908   & $4.58 \cdot 10^{-5}$  & $6.77 \cdot 10^{-3}$\\\cmidrule{2-10} 

& \multirow{4}{=}{\centering  MF (MFBD)} & \multirow{4}{=}{\centering  40 } & \multirow{4}{=}{\centering  [64, 128, 64, 32] } & \multirow{4}{=}{\centering  300 }  
        & \multirow{2}{*}{$\omega_{n}$}   & Train & 0.9951 & $9.14 \cdot 10^{-4}$ & $3.02 \cdot 10^{-2}$ \\
& & & & &                                 & Test  & 0.9919 & $1.65 \cdot 10^{-3}$ & $4.06 \cdot 10^{-2}$ \\\cmidrule{6-10}          
& & & & & \multirow{2}{*}{$\sigma_{max}$} & Train & 0.9061 & $3.31 \cdot 10^{-1}$ & $5.75 \cdot 10^{-1}$ \\ 
& & & & &                                 & Test  & 0.9049 & $2.60 \cdot 10^{-1}$ & $5.10 \cdot 10^{-1}$ \\
\Xhline{2pt}
\multirow{13}{=}{\centering  DWL } 
& \multirow{8}{=}{\centering  LF (MBD) } & \multirow{8}{=}{\centering  100 } & \multirow{8}{=}{\centering  [128, 64, 32, 16] } & \multirow{8}{=}{\centering  150 }  
        & \multirow{2}{*}{$P_{max}$}    & Train & 0.9856 & $9.11 \cdot 10^1$    & $3.02 \cdot 10^1$ \\
& & & & &                               & Test  & 0.9864 & $9.13 \cdot 10^1$    & $3.02 \cdot 10^1$ \\\cmidrule{6-10}        
& & & & & \multirow{2}{*}{$a_{max}$}    & Train & 0.9859 & $2.03 \cdot 10^5$    & $4.51 \cdot 10^2$ \\
& & & & &                               & Test  & 0.9865 & $2.01 \cdot 10^5$    & $4.49 \cdot 10^2$ \\\cmidrule{6-10} 
& & & & & \multirow{2}{*}{$\omega_{n}$} & Train & 0.9990 & $2.66 \cdot 10^{-4}$ & $1.63 \cdot 10^{-2}$ \\
& & & & &                               & Test  & 0.9987 & $3.07 \cdot 10^{-4}$ & $1.75 \cdot 10^{-2}$ \\\cmidrule{2-10}
& \multirow{4}{=}{\centering  MF (MFBD)} & \multirow{4}{=}{\centering  50 } & \multirow{4}{=}{\centering  [64, 128, 64, 32] } & \multirow{4}{=}{\centering  300 }  
        & \multirow{2}{*}{$\omega_{n}$}   & Train & 0.9962 & $1.81 \cdot 10^{-3}$ & $4.25 \cdot 10^{-2}$ \\
& & & & &                                 & Test  & 0.9913 & $1.91 \cdot 10^{-3}$ & $4.36 \cdot 10^{-2}$ \\\cmidrule{6-10}        
& & & & & \multirow{2}{*}{$\sigma_{max}$} & Train & 0.9950 & $1.16 \cdot 10^1$    & $1.08 \cdot 10^1$ \\ 
& & & & &                                 & Test  & 0.9862 & $1.71 \cdot 10^1$    & $1.31 \cdot 10^1$ \\
\Xhline{2pt}
\multirow{13}{=}{\centering  DWH } 
& \multirow{8}{=}{\centering  LF (MBD) } & \multirow{8}{=}{\centering  150 } & \multirow{8}{=}{\centering  [128, 64, 32, 16] } & \multirow{8}{=}{\centering  200 }  
        &\multirow{2}{*}{$P_{max}$}    & Train & 0.9950 & $5.45 \cdot 10^1$    & $2.33 \cdot 10^1$ \\
& & & & &                              & Test  & 0.9949 & $5.89 \cdot 10^1$    & $2.43 \cdot 10^1$ \\\cmidrule{6-10}           
& & & & &\multirow{2}{*}{$a_{max}$}    & Train & 0.9869 & $1.58 \cdot 10^5 $   & $3.98 \cdot 10^2$ \\
& & & & &                              & Test  & 0.9854 & $1.75 \cdot 10^5 $   & $4.19 \cdot 10^2$ \\\cmidrule{6-10}   
& & & & &\multirow{2}{*}{$\omega_{n}$} & Train & 0.9990 & $1.97 \cdot 10^{-4}$ & $1.41 \cdot 10^{-2}$\\
& & & & &                              & Test  & 0.9989 & $2.06 \cdot 10^{-4}$ & $1.43 \cdot 10^{-2}$ \\\cmidrule{2-10}
& \multirow{4}{=}{\centering  MF (MFBD)} & \multirow{4}{=}{\centering  30 } & \multirow{4}{=}{\centering  [128, 64, 32, 16, 8] } & \multirow{4}{=}{\centering  300 }  
        &\multirow{2}{*}{$\omega_{n}$}   & Train & 0.9982 & $5.64 \cdot 10^{-4}$ & $2.37 \cdot 10^{-2}$ \\
& & & & &                                & Test  & 0.9966 & $1.03 \cdot 10^{-3}$ & $3.20 \cdot 10^{-2}$ \\\cmidrule{6-10}           
& & & & &\multirow{2}{*}{$\sigma_{max}$} & Train & 0.9846 & $1.43 \cdot 10^1$    & $1.20 \cdot 10^1$ \\ 
& & & & &                                & Test  & 0.9962 & $8.22 \cdot 10^{-1}$ & $9.06 \cdot 10^{-1}$ \\
\Xhline{2pt}
\botrule
\end{tabular*}
\footnotetext{Note: Computational environment used for training both the LF and MF models was consistent across all systems. The models were trained using the Adam optimizer, the ReLU activation function, and a learning rate of 0.001. All computations were performed on an Intel(R) Xeon(R) CPU @ 2.00GHz.}
\end{table}



\begin{figure}[h!]
\centering
\includegraphics[width=1\textwidth]{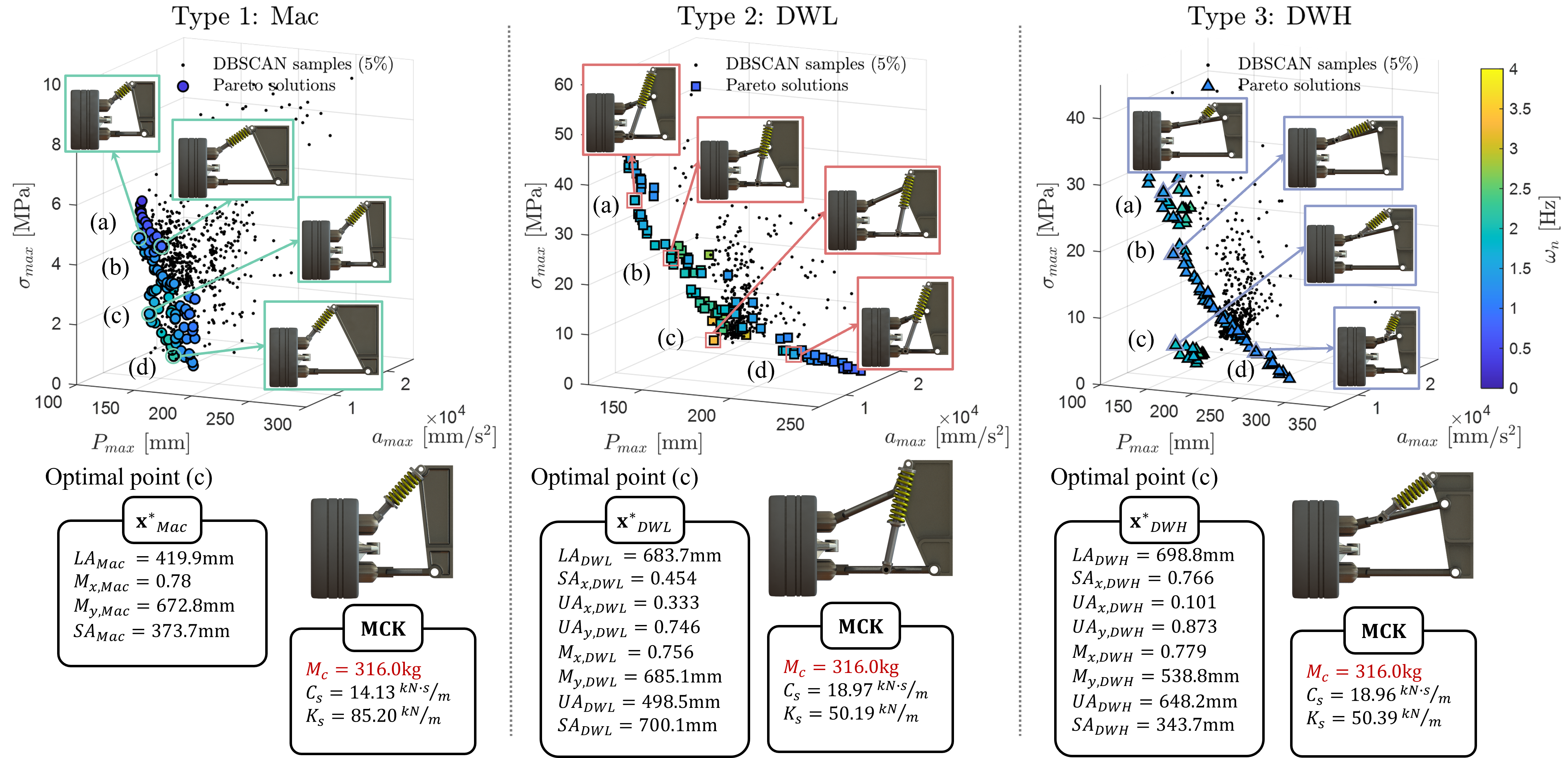}
\caption{Pareto solutions via NSGA-II  for the optimized ($P_{max}, a_{max}, \sigma_{max}$) three suspension types. The solutions are derived from the MF model trained by DBSCAN-sampled data.}
\label{fig:Pareto_3D}
\vspace{-9pt}
\end{figure}





\subsection{Optimization results}\label{subsec4_3}
\begin{figure}[h!]
\centering
 \includegraphics[width=1\textwidth]{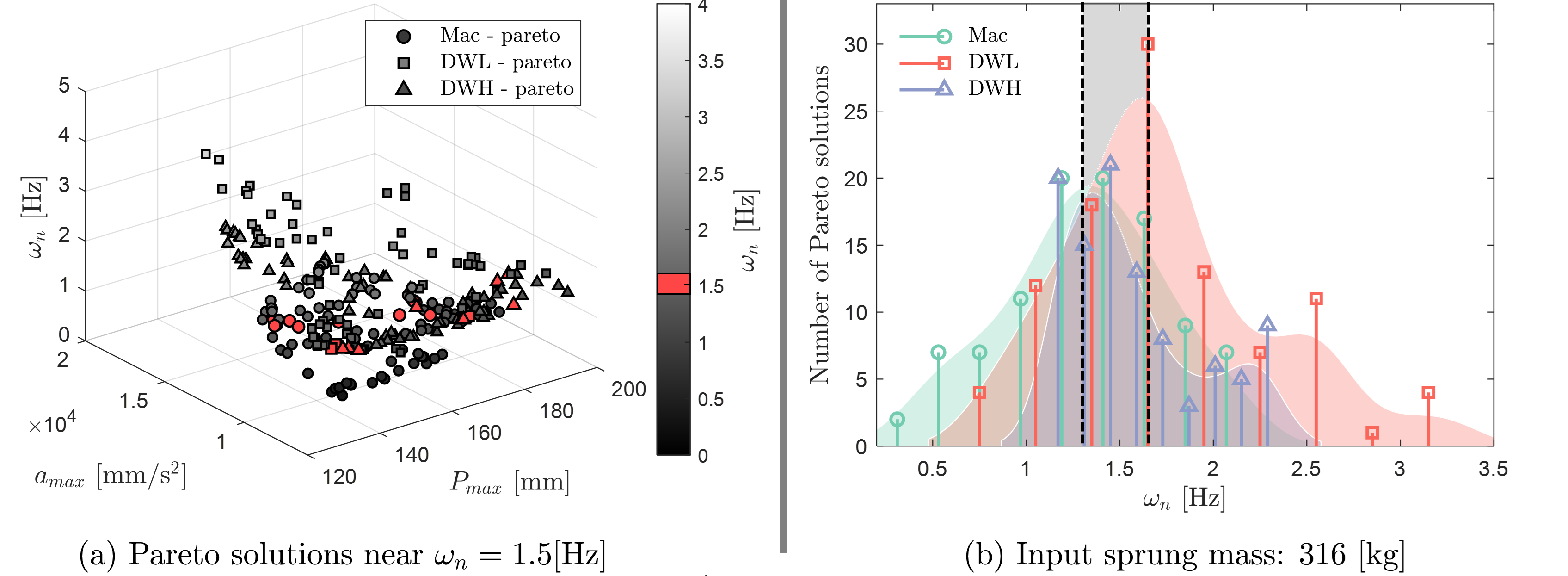}
\caption{(a) Optimized Pareto solutions for $P_{max}$ and $a_{max}$, with corresponding $\omega_{n}$ values represented through the color bar, allowing for an objective comparison across suspension types. (b) Distribution of near $\omega_{n}=1.5Hz$ values for the optimal candidates with an input sprung mass of $316 kg$, providing insight into the range of natural frequencies for informed selection of design variables.}
\label{fig:MF_Pareto_total}
\vspace{-9pt}
\end{figure}

The NSGA-II optimization was performed based on the trained MF models for each suspension type. The input parameter used was the sprung mass ($M_c$) set to $316 kg$, ensuring that it remained within the MCK conditions used during training. By identifying the optimized design variables and the corresponding $\omega_{n}$ for each suspension type with a single input sprung mass, we aim to provide designers with optimal candidates for selecting the appropriate type, design values, and reference measures. The optimization process yielded results for the design variables optimized for three performance metrics ($P_{max}, a_{max}, \sigma_{max}$). Fig.~\ref{fig:Pareto_3D} illustrates the Pareto solutions representing the optimal performance for each type. Comparing the samples generated by DBSCAN with the predicted model data samples confirms that the Pareto front was effectively identified at the optimal boundary. While displaying all Pareto solutions for the three performance metrics on the same axis would be ideal, the comparison is limited because $\sigma_{max}$ values correspond to different components applied to each suspension type, making an objective comparison challenging. Consequently, Fig.~\ref{fig:MF_Pareto_total} (a) presents an objective comparison using $P_{max}, a_{max}$ performance metrics along with their corresponding $\omega_{n}$ values, visualized using a color bar.

Though not an optimizable metric, $\omega_{n}$ value is a reference after the design process. As shown in Fig.~\ref{fig:MF_Pareto_total} (a), the optimized candidates for the input sprung mass of $316 kg$ exhibit a range of $\omega_{n}$ values. To better understand the distribution of these $\omega_{n}$ values, Fig.~\ref{fig:MF_Pareto_total} (b) provides a numerical distribution. This allows designers to select candidates with appropriate $\omega_{n}$ values, facilitating the selection of the optimal suspension type and design variables.


\begin{figure}[htb]
\centering
 \includegraphics[width=1\textwidth]{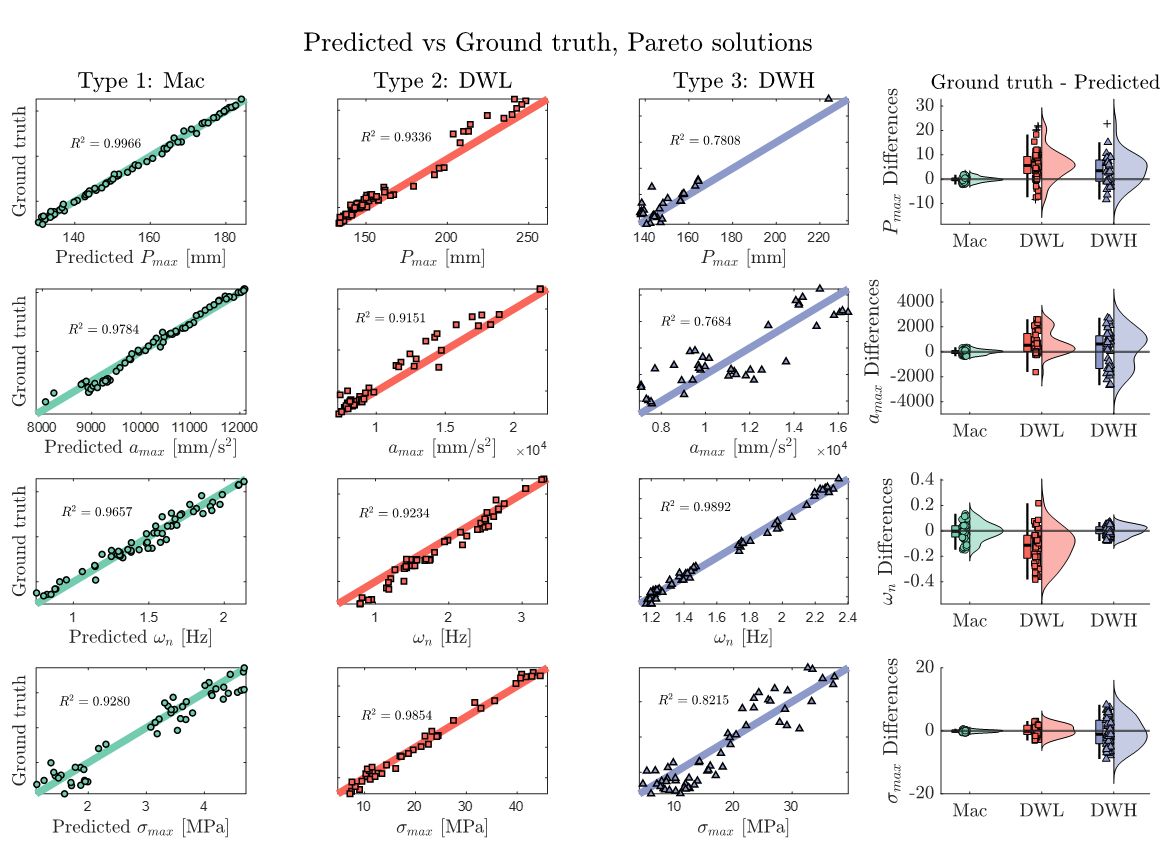}
\caption{Comparison of predicted performance values with ground truth obtained from MFBD analysis for each design type. The leftmost plots show the predicted versus actual values, with the coefficient of determination ($R^2$) indicating the prediction accuracy.}
\label{fig:Pred_vs_GT}
\vspace{-9pt}
\end{figure}

The optimal design variables and predicted performance values obtained from the Pareto solutions were evaluated against the actual results from the MFBD. The comparison between the predicted performance and the MFBD-derived ground truth for each design type is shown in Fig.~\ref{fig:Pred_vs_GT}.
Overall, the predicted performance values follow the general trend of the actual results. However, discrepancies were observed when comparing these results to the evaluation metrics of the samples used to train the prediction model. This divergence can be attributed to the fact that the optimal design variables obtained through NSGA-II were not part of the training data. These optimal points likely reside at the boundaries of the performance space, which were not fully explored during model training, leading to the observed differences.
Nevertheless, the results indicate that the predictions for $\omega_{n}$ are relatively accurate across the different types. This indicates that the value of $\omega_{n}$ can serve as a reliable indicator for the selection of appropriate types and design variables, based on the desired input sprung mass. This underscores the significance of $\omega_{n}$ in providing direction during the design process and in fulfilling the objective of achieving satisfactory performance results.

\subsection{Comparison pareto solutions: LF vs MF model frameworks}\label{subsec4_4}
\begin{figure}[h!]
\centering
 \includegraphics[width=1\textwidth]{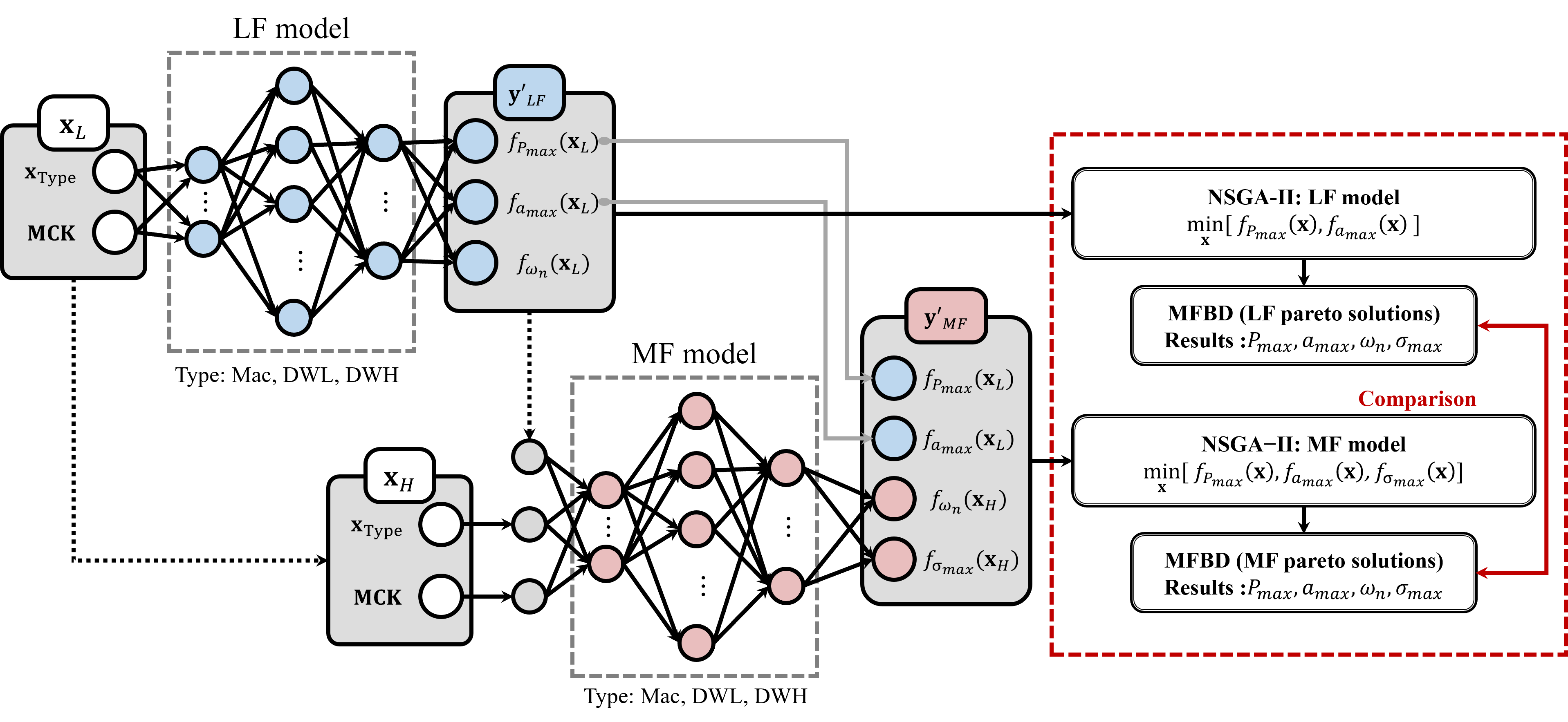}
\caption{This diagram shows the optimization and comparison process between LF and MF models. The LF model generates Pareto solutions by optimizing two performance metrics, $P_{max}$ and $a_{max}$. The MF model, on the other hand, optimizes three metrics: $P_{max}$, $a_{max}$, and $\sigma_{max}$ using additional performance data, including $\omega_n$. Each set of Pareto solutions is then subjected to MFBD simulations to obtain the final results. Then, performance metrics are compared to evaluate the effectiveness of the two models in predicting optimal design.}
\label{fig:Comp_LF_MF_diagram_1}
\vspace{-9pt}
\end{figure}

This section compares the efficiency of the conventional optimization process using prediction models with the proposed MF model framework. Prediction models effectively reduce computational costs and efficiently predict performance by utilizing sufficient data. However, in industrial applications, the performance metrics required at each analysis stage may differ as the process progresses, or entirely different performance metrics may need to be considered. In such cases, creating prediction models for each analysis stage to perform optimization can be impractical. An example is the suspension mechanism design, where significant differences in performance metrics between low-fidelity (MBD) and high-fidelity (MFBD) analyses were observed(in Fig.~\ref{fig:MBD_vs_MFBD}). Despite these challenges, this study compares the Pareto solutions obtained by performing NSGA-II using the LF model and MFBD analysis with those obtained using the MF model and MFBD analysis. Figure.~\ref{fig:Comp_LF_MF_diagram_1} illustrates the optimization process used to compare the two frameworks. For a fair comparison, the optimization process was performed under the same conditions, and the LF model only considered two performance metrics, $P_{max}$ and $a_{max}$, as the $\sigma'_{max}$ value cannot be obtained in MBD analysis.

\begin{figure}[h!]
\centering
 \includegraphics[width=1\textwidth]{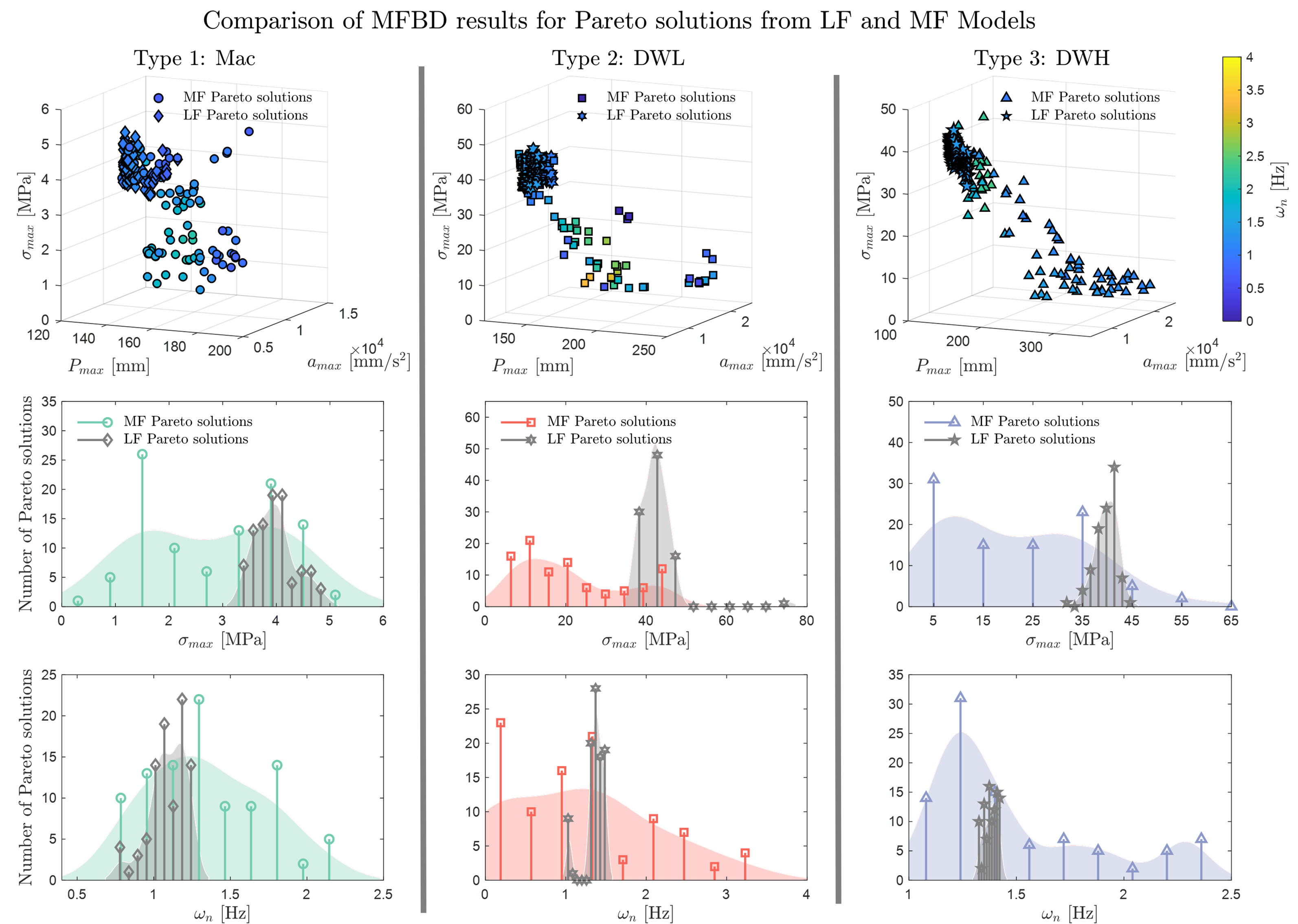}
\caption{Comparison of MFBD Simulation Results for Pareto Solutions from LF and MF Models. The 3D scatter plots (top and mid row) reveal that MF model solutions generally produce lower and more evenly distributed $\sigma_{max}$ values, while LF model solutions tend to show higher stress levels, implying greater stress on suspension links. Histograms (bottom row) indicate that MF solutions offer a wider range of $\omega_n$ values, allowing for more flexible design choices. Conversely, LF solutions are clustered within a narrow $\omega_n$ range, limiting the selection of optimal designs. This underscores the limitations of the LF model in providing diverse and robust design options.}
\label{fig:MF_GT_vs_LF_GT}
\vspace{-9pt}
\end{figure}

The results of each process are shown in Fig.~\ref{fig:MF_GT_vs_LF_GT}. First, a scatter plot was used to visualize the differences in all performance values for the Pareto solutions (top row). The mid-row of Fig.~\ref{fig:MF_GT_vs_LF_GT} shows the $\sigma_{max}$ values distribution for each framework. A significant difference was observed depending on whether $\sigma_{max}$ was included as an objective function, which aligns with the importance of considering MFBD performance. Therefore, the maximum stress values for the flexible body in the LF model were generally higher than those in the MF model's Pareto solutions, indicating a limitation in fully utilizing the design potential within the LF model. The bottom row of Fig.~\ref{fig:MF_GT_vs_LF_GT} shows the distribution of $\omega_{n}$ values for the Pareto solutions. Comparing the two frameworks, the $\omega_{n}$ values obtained from the MF model show a more uniform and broader distribution than those from the LF model. This suggests that, when selecting designs based on $\omega_{n}$ after performing optimization based on the input sprung mass, the LF model allows for a narrower range of choices, limiting the ability to fully explore the potential of design space. In contrast, the MF model's Pareto solutions demonstrate a relatively wider distribution, highlighting the effectiveness of the proposed MF model framework and its potential for effectively utilizing performance-based design in various applications. This broader distribution emphasizes the critical role of high-fidelity analyses in capturing nuanced performance characteristics, allowing for a more robust exploration of complex design trade-offs. As such, the multi-fidelity approach can significantly enhance the decision-making process in optimizing intricate engineering systems.

\subsection{Design Rule Extraction}\label{subsec4_4}
We conducted several additional analyses to extract design rules for the optimization of suspension mechanism using the trained deep learning model, which allows real-time evaluation of various design configurations. First, in Section \ref{sec:Sobol}, a sensitivity analysis using the Sobol method was carried out over the entire design space to identify the influence of each design parameter on the objective functions(\citeauthor{sobol2001global, herman2017salib}). Furthermore,in Section \ref{sec:PCA}, PCA was employed to analyze the principal components of the design variables in relation to the objective functions. Then, in Section \ref{sec:DT}, a decision tree algorithm was employed around the optimal design space to extract multivariate design rules. Lastly, in Section \ref{sec:Coef}, Spearman and Pearson correlation analyses were performed to further investigate the relationships between design parameters and performance outcomes near the optimal solutions. For further details on the data mining techniques applied in this study, please refer to the work(\citeauthor{yang2022design}).

\subsubsection{Sobol sensitivity analysis}\label{sec:Sobol}
\begin{figure}[h!]
\centering
 \includegraphics[width=1\textwidth]{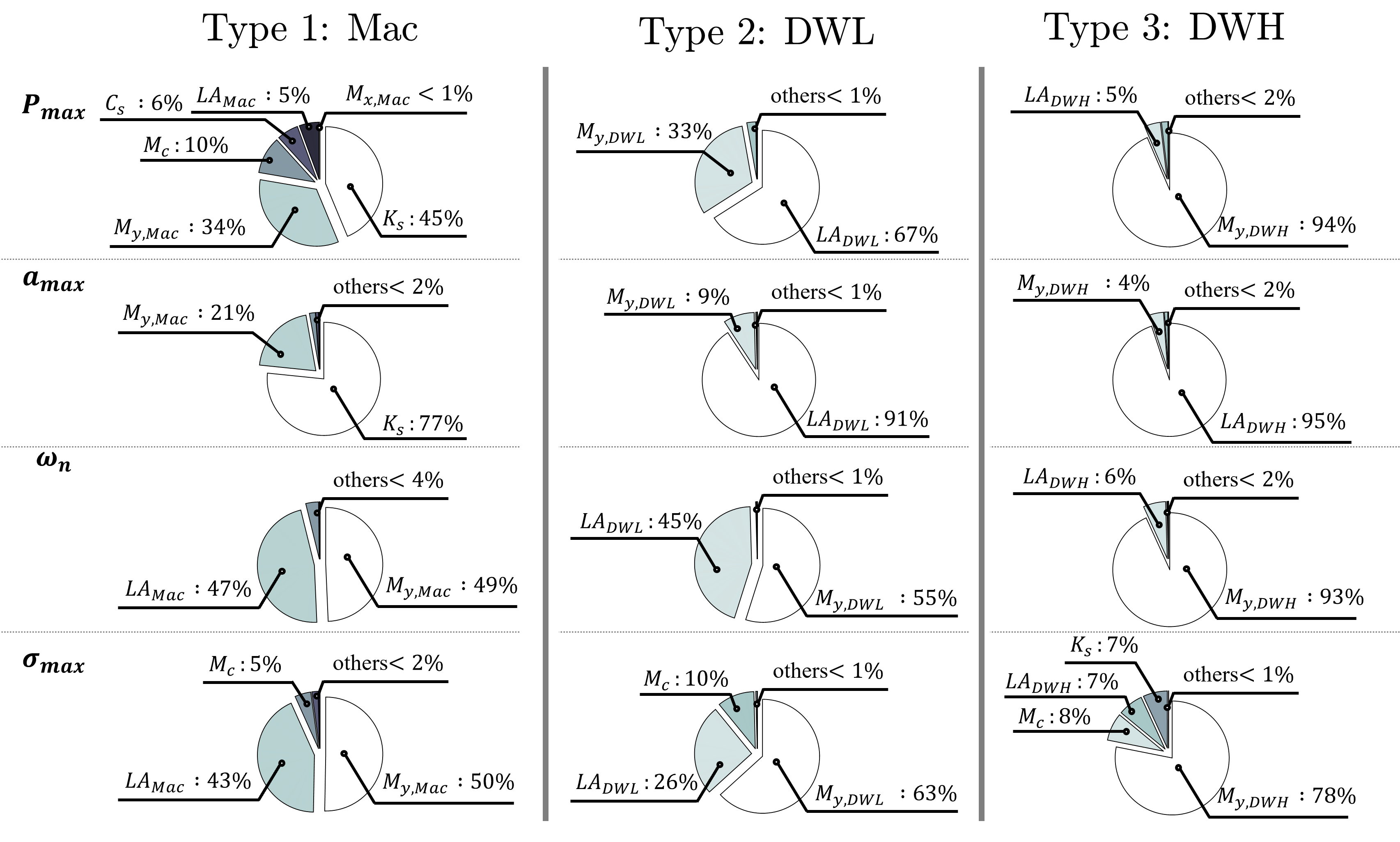}
\caption{Sobol results for each type and performances.}
\label{fig:Sobol}
\vspace{-9pt}
\end{figure}

A total of 300 (Mac), 209 (DWL), and 200 (DWH) design points were sampled for each suspension type using a Sobol sampler within the predefined design space. To this end, the Python library \textsc{SALib} was employed. The results of the Sobol sensitivity analysis for the performance metrics with respect to the design variables are shown in Fig. \ref{fig:Sobol}. For the Mac type, the contribution of the shock absorber spring stiffness ($K_s$) was found to be significant for both $P_{max}$ and $a_{max}$ values. This is likely due to the relatively simple structure of the Mac system, where fewer links are involved, making the shock absorber, which is directly mounted to the knuckle, have a more prominent impact. Additionally, the $\omega_{n}$ and $\sigma_{max}$ were significantly affected by the height of the mount ($M_{y,Mac}$). This is consistent with that the mount height determines the range of motion, directly affecting the system's natural frequency and the maximum stress the structure encounters.

For the DWL type, the $P_{max}$ and $a_{max}$ values were strongly influenced by the position of the lower arm where the shock absorber is mounted. This aligns with the general understanding that the mounting position of the shock absorber can significantly affect vehicle performance, especially under dynamic conditions. Similarly, for $\omega_{n}$ and $\sigma_{max}$, the height of the mount ($M_{y,DWL}$) was a key variable, again reflecting a similar pattern as seen in the Mac type.

For the DWH type, $P_{max}$ was primarily influenced by the mounting position of the upper shock absorber, making the mount height ($M_{y,DWH}$) a critical factor. On the other hand, $a_{max}$ was more sensitive to the $LA_{DWH}$ variable, which defines the maximum range of movement for the knuckle. As with the Mac and DWL types, the natural frequency and maximum stress were highly dependent on the mount height, underscoring the importance of this variable across all types for providing optimal dynamic performance.




\subsubsection{Principal component analysis (PCA)}\label{sec:PCA}
\begin{figure}[h!]
\centering
 \includegraphics[width=0.5\textwidth]{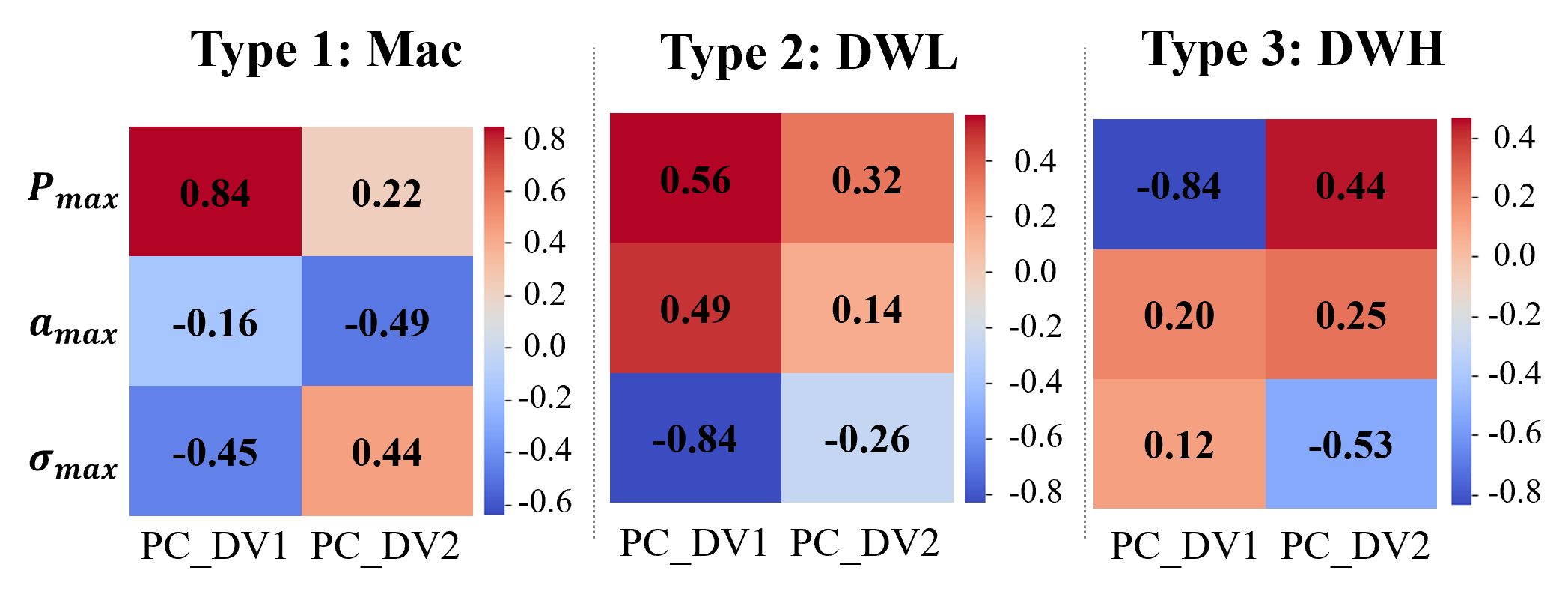}
\caption{PCA results for each type and objective functions.}
\label{fig:PCA}
\vspace{-9pt}
\end{figure}

Principal component analysis (PCA) was performed to identify the dataset's key patterns and reduce the design variables' dimensionality by transforming them into uncorrelated principal components (PCs). Each principal component is a linear combination of the original design variables, capturing the maximum possible variance in the data while minimizing information loss. This technique provides insight into how each design variable influences the overall system performance and highlights the interactions between them. The results of the PCA for the Mac, DWL, and DWH suspension types are displayed in Fig. \ref{fig:PCA}. Each plot shows the principal components and their importance, representing each design variable's weight in the corresponding principal component. The design variables are mapped into new variables ($PC\_DVs$), where each $PC\_DV$ is a weighted sum of the original design variables.

For the Mac suspension, the first principal component ($PC\_DV1$) accounts for $33\%$ of the variance in the design space. This component exhibits high negative loadings from $M_{y,Mac}$, and $K_s$, indicating a significant correlation among these variables and their similar influence on system performance. The second principal component ($PC\_DV2$), which explains $20\%$ of the variance, reveals strong positive correlations with $LA_{Mac}$ and $M_c$, further highlighting the critical roles of these variables in determining performance results.

In the DWL suspension, the first principal component ($PC\_DV1$) accounts for $28\%$ of the variance, with significant contributions from $LA_{DWL}$, and $SA_{x,DWL}$. This component dominates the design space, capturing the critical factors that govern the dynamic behavior of the suspension system. The second principal component ($PC\_DV2$) explains $15\%$ of the variance, with notable contributions from $UA_{x,DWL}$ and $M_{x,DWL}$, highlighting the importance of the shock absorber mounting position and the arm configurations.

For the DWH suspension, $PC\_DV1$ explains $26\%$ of the variance, with substantial loadings from $SA_{x,DWH}$, and $UA_{y,DWL}$. The second principal component ($PC\_DV2$) captures $12\%$ of the variance, with contributions from $K_{s}$. These components highlight the significant influence of these variables on performance.

\subsubsection{Decision tree}\label{sec:DT}
\begin{figure}[h!]
\centering
 \includegraphics[width=1\textwidth]{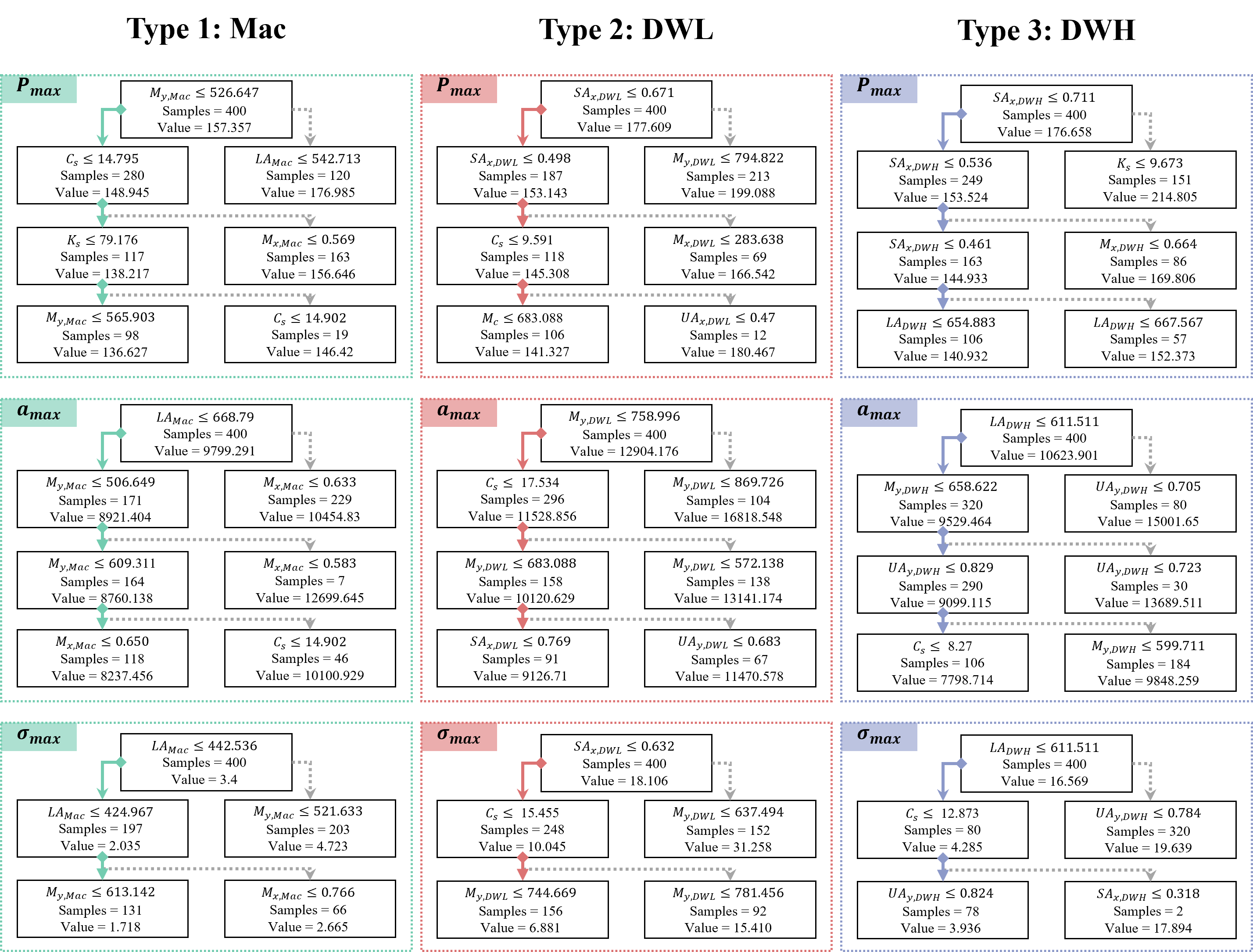}
\caption{Decision tree results for each type and objective functions.}
\label{fig:tree}
\vspace{-9pt}
\end{figure}

In contrast to the Sobol analysis mentioned in the previous subsection, this section aims to use decision tree analysis to explore the trade-offs between each type of suspension of the objective function. To achieve this, an additional 400 samples were drawn from the trained LF and MF models for each suspension type. Of the 400 samples, 300 were taken from the final three stages of the NSGA-II population, and 100 were selected from the Pareto solutions. The decision tree results are presented in Fig. \ref{fig:tree}, where the bold arrows highlight the optimization directions, indicating that all objective functions are minimized.

In the initial decision tree, the design variables with high influence on the objective functions were selected, as identified in the Sobol analysis through the $PC\_DV$. For instance, design elements with strong correlations to performance, such as the lower arm and mounting position, were prominently chosen. Furthermore, the decision tree guides how these design variables should be adjusted to optimize the objective functions, offering practical insights for making objective and informed decisions in the design process.


\subsubsection{Correlation coefficients}\label{sec:Coef}
\begin{figure}[h!]
\centering
 \includegraphics[width=1\textwidth]{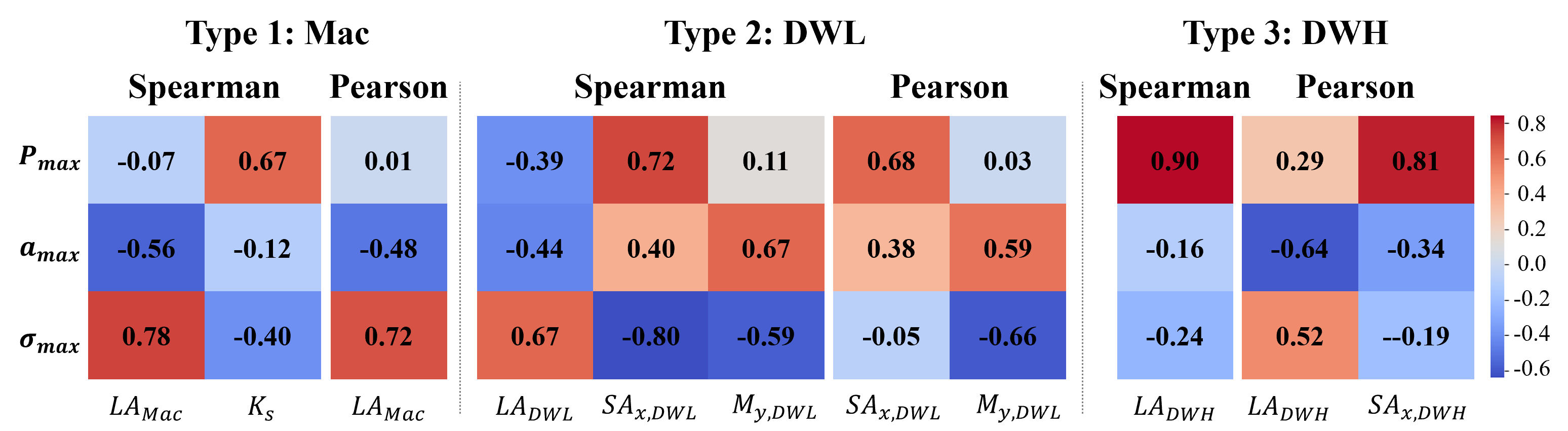}
\caption{Spearman and Pearson correlations between design variables and objective functions.}
\label{fig:Corr}
\vspace{-9pt}
\end{figure}

This subsection performs a correlation analysis to intuitively compare the design rules derived from the decision tree analysis. This analysis comprehensively explains the relationships between the input design variables and performance metrics. To achieve this, both Spearman and Pearson correlation coefficients were computed for the 400 design samples. The results, illustrated in Fig. \ref{fig:Corr}, reveal that key design variables, such as the lower arm and mounting position, mentioned in the Sobol analysis, consistently influence performance results.

The correlation analysis results for the Mac suspension type indicate that $LA_{Mac}$ positively correlates with $\sigma_{max}$. This suggests that, as the length of the lower arm increases, the range of motion for the knuckle also increases, leading to higher stress levels. This is a physically reasonable conclusion given the assumption of a frame with uniform thickness. Additionally, $K_s$ (spring stiffness) shows the second-highest correlation with $P_{max}$, reflecting that the structurally simpler Mac system has a greater dependency on the shock absorber. This finding is consistent with the Sobol analysis.

For the DWL suspension, $SA_{x,DWL}$ displays a strong positive correlation with $P_{max}$ and a strong negative correlation with $\sigma_{max}$. This indicates that as the mounting position of the shock absorber moves further from the knuckle, the maximum load increases, but the shock absorber becomes less effective in mitigating stress. Conversely, when the shock absorber is positioned closer to the mounting point, the moment arm of the lower arm is reduced, resulting in lower $\sigma_{max}$. This phenomenon can be physically explained by the reduced moment arm for rotation at the mounting position.

Lastly, in the DWH suspension type, a strong correlation was observed between $LA_{DWH}$ and both $P_{max}$ and $SA_{x,DWH}$. Since the shock absorber is mounted to the upper arm, geometrical changes to the lower arm can significantly influence the performance due to its dependency on related design variables. This finding highlights that the mechanical structure and characteristics of the components significantly affect the system's performance, underscoring the importance of optimizing the mechanism design for improved performance.





\section{Conclusion}\label{sec5}
This paper proposes a novel multi-fidelity optimization framework designed to address the challenges of optimizing different types of suspension mechanisms that perform the same function, focusing on vehicle suspensions. Traditional optimization approaches usually aim to maximize or minimize performance based on design variables. However, industrial field design processes are more complex, concerning constraints that cannot be optimized and requiring careful consideration of trade-offs across multiple stages. The proposed framework accommodates these complexities by including reference values for parameters that are difficult to optimize directly and balancing performance metrics using DBSCAN to navigate multi-fidelity challenges. The need for a robust, multi-stage framework in iterative design processes, common in the industry, is evident when different analysis conditions can lead to differences. A single predictive model may not fully capture the intricacies of such processes, necessitating a comprehensive approach like the one presented in this study.

The framework is built around the simultaneous optimization of multiple mechanism types, each with a distinct set of design variables. It facilitates the exploration of parametric design spaces and enables mechanism selection, especially when direct optimization of certain reference performance values is not feasible. By combining low-fidelity (LF) and high-fidelity (HF) models and leveraging both MBD and MFBD analyses, the framework effectively minimizes discrepancies between low-fidelity, less accurate simulations and high-fidelity, more precise evaluations.
The framework was validated by comparing Pareto solutions derived from LF models with those obtained using the proposed multi-fidelity framework. Results demonstrate that the MF approach provides more reliable predictions of optimal design solutions, particularly for complex systems where multiple performance criteria are involved. Furthermore, the application of data mining techniques, such as Sobol sensitivity analysis and decision tree methods, facilitated the extraction of practical design rules, highlighting the most influential design variables in system performance. These rules guide the designer in understanding trade-offs between competing objectives and optimizing system performance more effectively. The scalability was demonstrated by successfully optimizing three different suspension mechanisms. A parametric design process was employed to generate 3D CAD models for each mechanism, followed by MBD analyses. DBSCAN was then used to select samples for MFBD analysis, effectively reducing inconsistencies between the two fidelity levels. NSGA-II was applied to obtain Pareto-optimal solutions, with comparisons between LF and MF models showing that the LF models resulted in generally higher stress values, indicating more low-grade performance and a much narrower range of natural frequencies, limiting the design flexibility. These findings underscore the superiority of the MF model in providing more accurate and comprehensive evaluations, highlighting the importance of incorporating both high- and low-fidelity analyses to ensure a more robust understanding of the design space.

The key contributions of this study are as follows: to the best of our knowledge, this is the first framework that optimizes multiple types of mechanisms while considering reference performance values that are challenging to optimize. The framework also minimizes discrepancies between low-fidelity and high-fidelity analyses, providing a more reliable and objective means of selecting and optimizing mechanisms that do not share common design variables.
This study presents a scalable, adaptable, and comprehensive framework for optimizing vehicle suspension mechanisms using multi-fidelity modeling and data-driven techniques. The methodology is not limited to suspensions and can be extended to other complex systems with similar design challenges, offering significant potential for broader applications in various engineering domains.

\backmatter

\bmhead{Acknowledgments}
This work was supported by the National Research Foundation of Korea grant (2018R1A5A7025409), the Ministry of Science and ICT of Korea grant (No.2022-0-00969, No.2022-0-00986), and the Ministry of Trade, Industry \& Energy (RS-2024-00410810).

\section*{Declarations}
\bmhead{Conflict of interest}
The authors declare that they have no known competing financial interests or personal relationships that could influence the work reported herein.

\bmhead{Replication of results}
The code and data are available from the corresponding author on reasonable request.









\bibliography{sn-bibliography}

\end{document}